\documentclass[sigconf]{acmart}
\usepackage{array}
\usepackage{booktabs}
\usepackage{multirow}
\usepackage[ruled,vlined,linesnumbered]{algorithm2e} 
\usepackage{subcaption}

\usepackage{tikz}
\usepackage{amsmath}

\usepackage{filecontents}

\def\A{\text{ATT\&CK}\xspace}
\def\S{\textit{TRACE}\xspace}

\AtBeginDocument{%
  \providecommand\BibTeX{{%
    \normalfont B\kern-0.5em{\scshape i\kern-0.25em b}\kern-0.8em\TeX}}}

\renewcommand\footnotetextcopyrightpermission[1]{}

\begin{document}

\title{CVE-to-ATT\&CK Mapping via Evidence-Constrained Ranking}

\author{Kai Wei}
\affiliation{
    \institution{University of South Florida}
    \city{Tampa}
    \state{FL}
    \country{USA}
}

\author{Yuwen Cui}
\affiliation{
    \institution{University of South Florida}
    \city{Tampa}
    \state{FL}
    \country{USA}
}

\author{Kehan Shen}
\affiliation{
    \institution{University of South Florida}
    \city{Tampa}
    \state{FL}
    \country{USA}
}

\author{Guangjing Wang}
\affiliation{
    \institution{University of South Florida}
    \city{Tampa}
    \state{FL}
    \country{USA}
}



\begin{abstract}
Mapping CVEs to MITRE ATT\&CK techniques helps defenders translate vulnerability information into hypotheses about adversary behavior. 
Automating this mapping is challenging because relevant evidence is fragmented across heterogeneous security sources. 
In addition, ATT\&CK techniques can be semantically similar, and existing annotations cover a highly long-tailed subset of the technique space. Existing supervised, retrieval-based, and LLM-based approaches therefore face limitations from closed-set supervision, semantic ambiguity, or unconstrained prediction.

We formulate CVE-to-ATT\&CK mapping as multi-evidence constrained ranking. Our framework TRACE first integrates complementary structural and semantic evidence from multiple sources to construct a bounded, high-coverage candidate pool over the complete ATT\&CK technique space. It then reranks the retained candidates by combining their accumulated evidence with CVE--technique textual interactions and frequency-aware scoring. We evaluate the framework across six public datasets with different annotation processes and vulnerability distributions. Candidate construction reduces the 201-technique ATT\&CK search space by approximately 60\% while retaining at least one ground-truth technique for 95.11\%--100\% of CVEs. These results show that combining heterogeneous security evidence with constrained ranking provides a more controlled and generalizable bridge from vulnerability information to adversary behavior.

\end{abstract}

\maketitle

\fancyhead[LE,RO]{}

\section{INTRODUCTION}
The number of reported vulnerabilities grows rapidly~\cite{nist2026nvd}.
The reasons are twofold.
First, the volume of code surges with agentic AI coding, which expands the vulnerability surface~\cite{guan2026aiagentsenableadaptive, pearce2021asleepkeyboardassessingsecurity, haque2025sokexploringhallucinationssecurity}. Second, many AI-driven security tools discover new vulnerabilities at scale. For example, Claude Mythos Preview identified thousands of vulnerabilities across major operating systems and web browsers~\cite{anthropic_mythos_preview}.

A vulnerability report tells defenders which software flaws are present.
For example, identified vulnerabilities are reported to the Common Vulnerabilities and Exposures (CVE) system.
Each CVE record typically describes an affected product, exploitation conditions, and reported consequences. Yet, a CVE does not provide a complete account of how an adversary may operationalize the vulnerability within an attack.
Thus, identifying a vulnerability is only the first step toward understanding its security implications. 

To understand potential adversary behaviors, MITRE ATT\&CK\footnote{Enterprise ATT\&CK comprises 201 parent-level techniques, while some of the techniques are divided into sub-techniques. Version: v14.1} provides a standardized framework that categorizes adversarial objectives into tactics and the methods used to achieve objectives into techniques~\cite{mitre_attack}.
Mapping CVEs to \A techniques establishes a connection between software vulnerabilities and adversarial behaviors: CVEs identify specific software flaws, while \A techniques describe the actions attackers may perform by exploiting these vulnerabilities to achieve their objectives. 
Such mappings provide evidence-constrained behavioral hypotheses that analysts can combine with asset context and threat intelligence to reason about how a vulnerability may contribute to an attack.

Different methods are proposed to achieve CVE-to-\A mapping. 
For example, supervised classifiers~\cite{ampel2021linkingcommonvulnerabilitiesexposures, a15090314,info15040214, 10900988} learn direct CVE-to-technique associations but depend on the coverage and frequency of labels in the training dataset. 
Knowledge-graph methods~\cite{hogan2021knowledge, bastos2021recon, hemberg2021linkingthreattacticstechniques} exploit curated relationships among security entities but cannot recover a relevant technique when the required graph path is missing. 
Semantic retrieval methods~\cite{guo2016deep, abdeen2023smet, li2025matching} can search a broader \A space. However, semantic similarity alone cannot distinguish techniques supported by concrete vulnerability evidence from those that merely share similar terminology.
More recent LLM-based approaches~\cite{liu-etal-2023-end,math12091286,yu2026vdm} provide flexible interpretation, but open-ended generation makes the prediction space difficult to constrain and may produce techniques that are insufficiently supported by the available evidence.

\begin{figure}[t]
\centering
\includegraphics[width=0.35\textwidth]{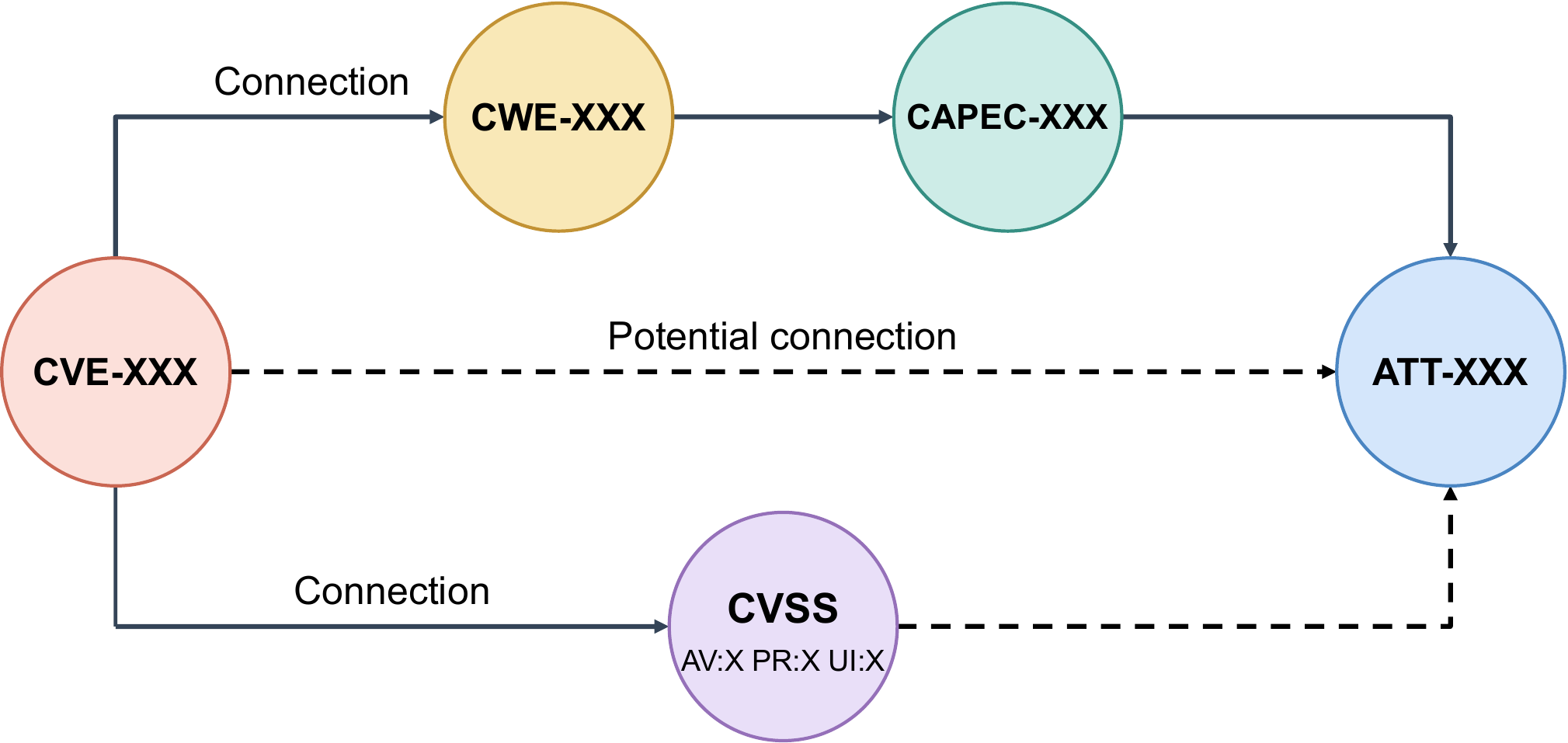}
\caption{The connectivity among security report sources.}
\label{fig:evidence}
\vspace{-10pt}
\end{figure}

Reliable CVE-to-\A mapping requires integrating evidence from heterogeneous security sources (datasets) to establish how a software vulnerability relates to specific adversarial behaviors.
For instance, CVE descriptions capture affected components and reported effects~\cite{cve}; CWE characterizes the underlying weakness~\cite{cwe}; CAPEC describes related attack patterns~\cite{capec}; 
CVSS encodes exploitability and impact conditions~\cite{cvss_nvd}. 
\figurename~\ref{fig:evidence} illustrates the relationships among CVEs, CWEs, CAPEC attack patterns, CVSS attributes, and MITRE ATT\&CK techniques, which connect vulnerabilities to adversarial behaviors.

Yet, each source in \figurename~\ref{fig:evidence} captures only part of the relationship between a vulnerability and adversary behaviors.
For example, the CVE-2021-44228 description reports \textit{JNDI-based arbitrary code execution}~\cite{nvd_cve_2021_44228}, CWE-917 identifies the underlying expression-language injection weakness~\cite{mitre_cwe_917}, and CVSS indicates network-based exploitation with low complexity, no required privileges, and no user interaction~\cite{nvd_cvss_2021_44228}. None of these sources explicitly identifies an \A technique. Nevertheless, taken together with the affected application's exposure context, T1190 \textit{Exploit Public-Facing Application} can be a plausible candidate.

In addition, we observe that retrieving candidates from multiple sources can yield several plausible \A techniques for the same CVE. This creates ambiguity about which techniques are supported by vulnerability evidence. 
For example, direct semantic retrieval also identifies T1059 (\textit{Command and Scripting Interpreter}) because "arbitrary code execution" in the CVE-2021-44228 description resembles "executing arbitrary commands" in T1059. 
However, arbitrary code execution does not necessarily involve a command or scripting interpreter, which makes additional evidence necessary to distinguish valid mappings from semantically plausible but unsupported candidates.

Based on the above observations, we formulate CVE-to-\A mapping as a heterogeneous data-integration problem.
Given a vulnerability, the task is to integrate structured and textual evidence to identify plausible candidates from the \A technique space. 
Constructing and evaluating candidate CVE-to-\A mappings presents two key challenges.

\begin{figure}[t]
\centering
\includegraphics[width=0.98\linewidth]{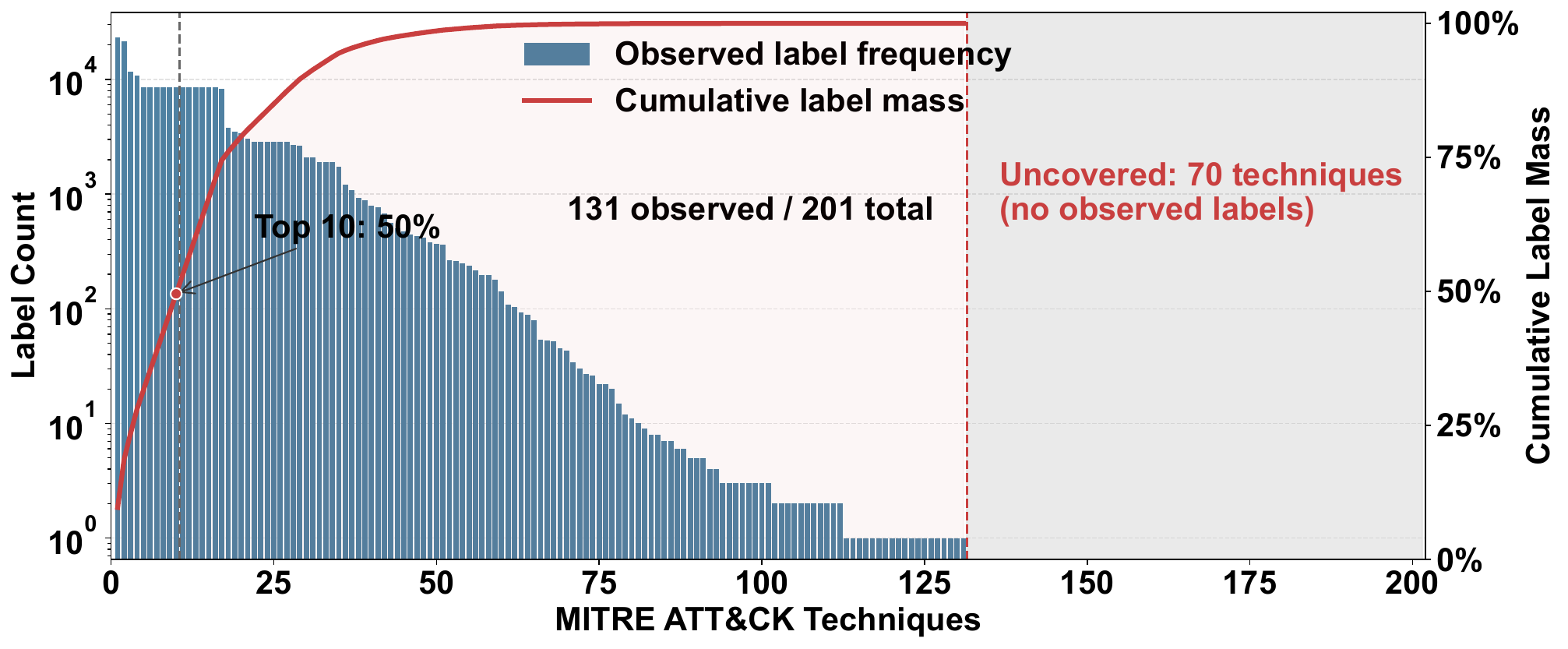}
\caption{
The label-frequency distribution over the complete ATT\&CK technique space.
}
\label{fig:tech_long_tail}
\end{figure}

\textbf{First, incomplete and imbalanced evidence limits candidate coverage.}
Security sources are developed for different purposes and through different annotation processes, resulting in incomplete and highly imbalanced technique coverage. As shown in \figurename~\ref{fig:tech_long_tail}, only 131 parent-level techniques appear across the six datasets considered in this study, with the ten most frequent techniques accounting for 50\% of all observed labels. Consequently, relevant techniques may be overlooked when supporting evidence is missing or indirect.
In addition, Table~\ref{tab:evidence-coverage} shows substantial variation in evidence availability across different security sources. For example, only 26.35\% of CVE records in the KEV dataset have an available CWE--CAPEC connection, highlighting the limitations of relying on a single evidence source and motivating complementary evidence paths.
\textbf{Second, multiple plausible candidates create mapping ambiguity.}
Different security sources may suggest different \A techniques for the same CVE, making it difficult to distinguish mappings from semantically plausible but weakly supported candidates. Ranking these candidates often relies on historical CVE-to-\A annotations, whose sparsity and imbalance further complicate reliable candidate selection.

\begin{table}[t]
\centering
\caption{Availability of structured evidence and CWE--CAPEC connectivity among CVE identifiers.}
\label{tab:evidence-coverage}
\scriptsize
\setlength{\tabcolsep}{3.5pt}
\begin{tabular}{@{}lrrrrr@{}}
\toprule
Dataset & CVEs & CWE & CVSS & CWE--CAPEC & All Evidence \\
\midrule
KEV~\cite{kev_mappings_explorer}
& 296 & 99 (33.45\%) & 262 (88.51\%)
& 78 (26.35\%) & 71 (23.99\%) \\
ATCM~\cite{attack_to_cve_mappings}
& 374 & 296 (79.14\%) & 339 (90.64\%)
& 253 (67.65\%) & 230 (61.50\%) \\
ENISA~\cite{enisa_vuln_report}
& 8,019 & 5,805 (72.39\%) & 7,978 (99.49\%)
& 5,391 (67.23\%) & 5,366 (66.92\%) \\
CVE2Attack~\cite{a15090314}
& 1,661 & 1,287 (77.48\%) & 1,486 (89.46\%)
& 994 (59.84\%) & 910 (54.79\%) \\
SMET~\cite{abdeen2023smet}
& 303 & 233 (76.90\%) & 280 (92.41\%)
& 192 (63.37\%) & 182 (60.07\%) \\
CGVM~\cite{liu-etal-2023-end}
& 31,298 & 28,570 (91.29\%) & 23,311 (74.48\%)
& 28,123 (89.86\%) & 20,785 (66.41\%) \\
\bottomrule
\end{tabular}
\vspace{-10pt}
\end{table}

In this work, we present \textbf{\S}, \textbf{T}echnique \textbf{R}anking with \textbf{A}ggregated \textbf{C}VE \textbf{E}vidence. \S is a multi-evidence constrained-ranking framework for mapping CVEs to MITRE ATT\&CK techniques.
The central principle is that ATT\&CK techniques should be prioritized according to the combined security evidence.
We first aggregate complementary structural and semantic evidence from different sources. 
The aggregation constructs a bounded, high-coverage candidate pool over the complete ATT\&CK technique space.
We then separately rerank the retained candidates by combining this accumulated evidence with CVE–technique textual interactions. 
This design allows candidate construction to optimize for coverage, while reranking optimizes for precision.

We evaluate \S across six public CVE-to-ATT\&CK datasets spanning different annotation processes and vulnerability distributions. 
Step~1 reduces the 201-technique ATT\&CK search space by 60\% while retaining at least one ground-truth technique for 95.11\%--100\% of CVEs, which provides a compact yet high-coverage search space for downstream ranking. 
Step~2 improves the ordering of these retained candidates, with NDCG@10 gains ranging from 0.0607 to 0.2806 across the six datasets. 
Against prior CVE-to-ATT\&CK mapping systems~\cite{abdeen2023smet,host2025systematicapproachpredictimpact}, TRACE achieves the strongest performance across all five ranking metrics on both KEV~\cite{kev_mappings_explorer} and CVE2Attack~\cite{a15090314} datasets. 
On CVE2Attack, for example, TRACE improves MRR from 0.4177 to 0.4319 and Hit@10 from 0.7261 to 0.7457 over the state-of-the-art TRIAGE~\cite{host2025systematicapproachpredictimpact}.
These results show that \S generalize beyond the datasets used for model development.
Overall, our main contributions are listed as follows:

\begin{itemize}

\item We formulate CVE-to-\A mapping as an evidence-constrained data integration problem. We recover relevant techniques and distinguish evidence-supported mappings from incomplete and heterogeneous security evidence.

\item We propose TRACE, a two-stage framework that integrates heterogeneous structural and semantic evidence to construct high-coverage candidate pools, then reranks techniques using evidence-aware and frequency-aware scoring.

\item We evaluate the \S across six public datasets. Step~1 reduces the ATT\&CK search space by 60\% while retaining at least one ground-truth technique for 95.11\%--100\% of CVEs, and Step~2 consistently improves ranking quality.

\end{itemize}

\begin{figure*}[t]
\centering
\includegraphics[width=\textwidth]{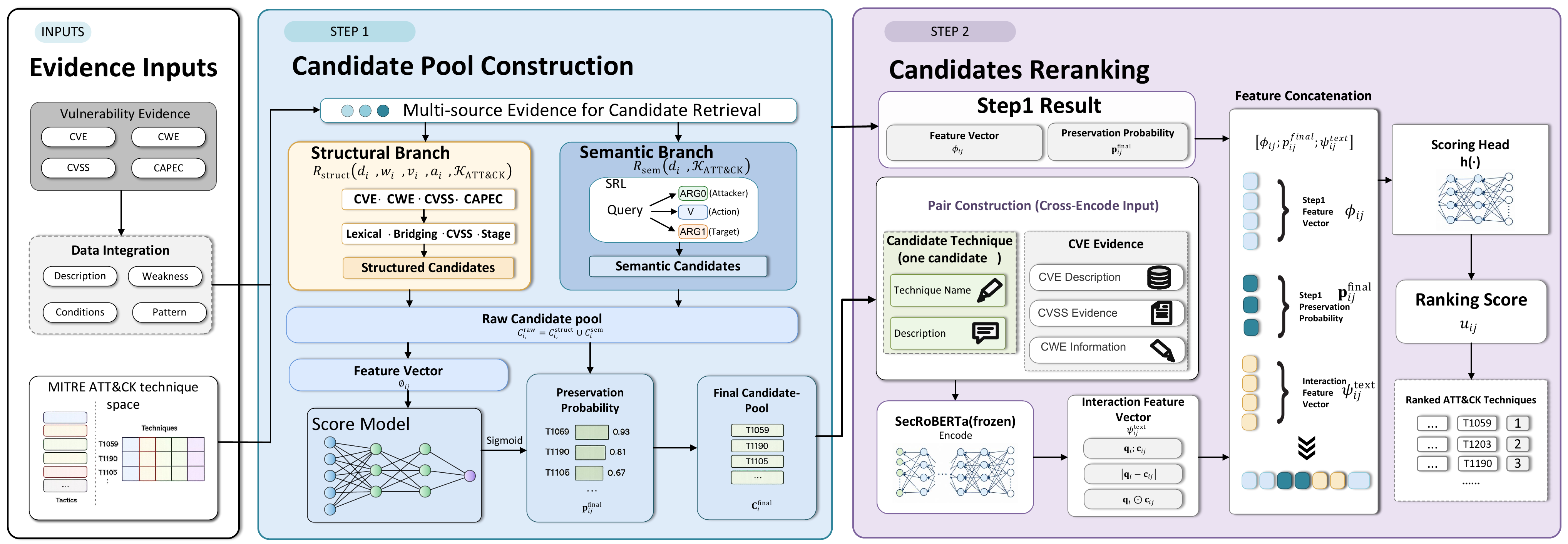}
\caption{Overview of the proposed two-step CVE to ATT\&CK mapping framework \S. Step 1 performs multi-source candidate recall and adaptive candidate preservation to build a high-coverage candidate pool, while Step 2 performs reranking over that candidate pool to produce the final \A techniques ranking. 
}
\label{fig1}
\end{figure*}

\section{RELATED WORK}
In this section, we introduce existing studies on CVE-to-ATT\&CK mapping, which can be broadly grouped into three categories.

\textbf{Knowledge-integrated mapping.}
The first category uses structured security knowledge to interpret vulnerability text. For example, BRON~\cite{hemberg2021linkingthreattacticstechniques} links CVE, CWE, CAPEC, and ATT\&CK records into a single relational graph.
This graph-based approach returns the paths available in the graph. But the path coverage is limited by the completeness of the curated CVE--CWE--CAPEC--ATT\&CK links.
VDM-IoT~\cite{11372793} builds on a similar idea by retrieving knowledge-graph paths to steer an LLM toward the correct mapping. 
But the knowledge-graph paths are supplied only as prompt context and do not constrain the LLM's output space. 
Consequently, the model may generate techniques unsupported by the retrieved evidence or omit relevant techniques whose supporting paths are absent.
These methods demonstrate the value of structured security knowledge, but they rely heavily on the availability of curated graph paths. When no path connects a CVE to a relevant technique, neither method provides a mechanism for recovering the missing connection. 
By contrast, our framework \S does not rely on a single curated knowledge-graph path to retrieve a candidate. Instead, it derives CVE--technique evidence from multiple complementary structural and semantic retrieval strategies. 
As a result, when a particular knowledge-graph relation is missing, a relevant technique can still be recovered through alternative evidence sources.

\textbf{Supervised and retrieval-based mapping.}
The second category models CVE-to-ATT\&CK mapping as multi-label classification. For example, CVE2ATTCK~\cite{a15090314} evaluates classical and BERT-based classifiers with data augmentation. 
Although these methods learn direct vulnerability-to-technique associations, they remain sensitive to training data distribution. If a technique does not appear in the training data, the classifier struggles to predict it correctly.

By contrast, retrieval-based methods such as SMET~\cite{abdeen2023smet} extract attack-related phrases from CVE descriptions and rank ATT\&CK techniques based on embedding similarity. However, semantic similarity alone may be insufficient when multiple techniques use similar terminology or describe related attack procedures. This is because their similarity scores may not reflect how strongly each technique is supported by the available vulnerability evidence. To address this limitation, \S jointly considers CVE--technique textual relations, candidate evidence features, and preservation probabilities, enabling it to prioritize strongly supported techniques over alternatives that are merely semantically similar.

\textbf{LLM-based and generation-based mapping.}
Recent studies apply LLMs directly to CVE-to-ATT\&CK mapping~\cite{liu-etal-2023-end,Rafiey2025MappingVD, math12091286,yu2026vdm}. TRIAGE~\cite{host2025systematicapproachpredictimpact}, for example, uses rule-guided prompting and in-context examples to steer LLM-based CVE-to-ATT\&CK mapping. Although such methods offer flexible semantic interpretation, their open-ended generation makes the prediction space difficult to constrain. As a result, they may produce redundant or semantically overlapping predictions, or techniques that are only weakly supported by the available vulnerability evidence.

\section{SYSTEM DESIGN}
\subsection{System Overview}
\label{sec:system_overview}

\S maps a CVE to a ranked list of ATT\&CK techniques through a two-step architecture as shown in Figure~\ref{fig1}. 
The design separates two objectives that are difficult to satisfy simultaneously: \emph{candidate coverage} and \emph{ranking precision}. 
A retrieval mechanism that is sufficiently permissive to preserve relevant and long-tail techniques inevitably introduces weakly related candidates, whereas an aggressive predictor risks discarding relevant techniques before they can be considered. 
Therefore, \S first constructs a bounded, high-coverage candidate pool and then performs fine-grained evidence-based reranking over the retained techniques.
 
\textbf{Step~1: Evidence-guided candidate construction.}
Given a CVE record $x_i$ containing its description, we incorporate auxiliary information including CWE information, CVSS attributes, and available CAPEC relations.
Step~1 retrieves candidates from the complete set of 201 parent-level Enterprise ATT\&CK techniques.
We design two complementary retrieval branches.
The \emph{structural branch} exploits direct lexical evidence, CVE--CWE--CAPEC--ATT\&CK relations, and CVSS-derived exploitation cues.
The \emph{semantic branch} extracts behavioral phrases from the CVE description using semantic role labeling and retrieves techniques according to semantic similarity to these behaviors.

However, as broad retrieval improves coverage but can introduce noisy candidates, \S records the evidence supporting each retrieved technique and uses a lightweight candidate-preservation model to estimate whether it should be retained.
The candidate pools are deduplicated into a single CVE-level pool.
The model then truncates the raw candidate set according to the accumulated probability mass, which produces a compact candidate pool $C_i^{\mathrm{final}}$.

\textbf{Step~2: Evidence-aware candidate reranking.}
The retained candidates in Step~1 may have similar preservation probabilities despite differing substantially in how well they are supported by the complete vulnerability evidence.
Therefore, Step~2 reranks $C_i^{\mathrm{final}}$ by jointly considering three signals: (i) the evidence features accumulated in Step~1, (ii) the candidate-preservation probability, and (iii) the semantic interaction between the CVE and the candidate ATT\&CK technique.
Specifically, we use SecRoBERTa~\cite{secroberta} to encode the CVE context and each candidate technique, and we fine-tune the scoring head of SecRoBERTa that combines their textual interaction features with the Step~1 evidence.
A frequency-based correction further reduces systematic preference toward frequently observed techniques.
The resulting scores are used to rank all retained candidates, which yield the final ATT\&CK prediction $\eta_i^{\mathrm{final}}$.

Overall, Step~1 favors recall so that relevant techniques are not irreversibly removed from consideration, while Step~2 favors precise ordering among the retained candidates.
\S does not rely on any single lexical match, knowledge-graph path, or semantic similarity score to determine the final prediction.
Instead, it aggregates complementary security evidence during candidate construction and preserves that evidence for the final ranking decision. 

\begin{algorithm}[t]
\caption{Step~1: Candidate-Pool Construction}
\label{alg:step1}
\DontPrintSemicolon
\small

\KwIn{Target CVE record $x_i=(d_i,w_i,v_i,a_i)$; ATT\&CK technique set $K_{ATT\&CK}$; Trained candidate-preservation model $F_\theta^{\mathrm{final}}$;  Score-mass target $\tau$ and Maximum pool size $K_{max}$.}

\KwOut{Final candidate pool $C_i^{\mathrm{final}}$; candidate-preservation scores $p_{ij}^{\mathrm{final}}$; and candidate feature vectors $\phi_{ij}$.}

$C_{i}^{\mathrm{struct}} \leftarrow  R_{\mathrm{struct}}(x_i,K_{ATT\&CK})$\;

$(C_i^{\mathrm{sem}},\{z_{ij}\})\leftarrow R_{\mathrm{\mathrm{sem}}}(d_i,K_{ATT\&CK})$\;

$C_{i}^{\mathrm{raw}}\leftarrow C_{i}^{\mathrm{struct}}\cup C_i^{\mathrm{sem}}$\;

\ForEach{$c_{ij}\in C_{i}^{\mathrm{raw}}$}{
$\phi_{ij}\leftarrow \operatorname{ConstructFeatures} \left(x_i, c_{ij}, K_{ATT\&CK}\right)$\;

$p_{ij}^{\mathrm{final}}\leftarrow \operatorname{Sigmoid}\!\left(F_\theta^{\mathrm{final}}(\phi_{ij})\right)$\; 
}

Sort $C_{i}^{\mathrm{raw}}$ in descending order of $p_{ij}^{\mathrm{final}}$\;

$C_{i}^{\mathrm{final}}
\leftarrow$ top-ranked candidates until cumulative normalized probability $\ge \tau$, with $|C_{i}^{\mathrm{final}}|\le K_{\max}$\;

\textbf{return} $C_i^{\mathrm{final}}$, $p_{ij}^{\mathrm{final}}$, $\phi_{ij}$

\end{algorithm}

\subsection{Evidence-Guided Candidate Construction}

We introduce Step~1 of \S to construct a compact, high-coverage ATT\&CK technique candidate pool for each CVE, as illustrated in Algorithm~\ref{alg:step1}.
Formally, given a CVE record $x_i$, where $x_i$ comprises the available CVE description $d_i$, CWE labels $w_i$, CVSS attributes $v_i$, and CAPEC relations $a_i$. 
We design two complementary retrieval branches, structural retrieval $R_{\mathrm{struct}}$ and semantic retrieval $R_{\mathrm{sem}}$, to retrieve candidate techniques using evidence from $x_i$.
The retrieval space consists of ATT\&CK techniques, denoted as $K_{ATT\&CK}$.

\subsubsection{Candidate Construction in Structural Branch}

In the structural retrieval branch $R_{\mathrm{struct}}$, the structural candidate set $C_{i}^{\mathrm{struct}}$ is constructed for each CVE. 
For each CVE, the framework applies three complementary retrieval strategies over the CVE record $x_i$ and the full ATT\&CK technique space $K_{ATT\&CK}$. 
The applicability of these three strategies varies depending on the availability of security evidence and the characteristics of individual CVEs.

First, \emph{direct lexical retrieval} compares the CVE and CWE description with each ATT\&CK technique's name and description, and retains a technique when they share at least one relevant non-stopword token. 
Parent-level lexical matching is also performed against parent-technique names and descriptions, allowing a parent technique to be recovered when the strongest lexical evidence is associated with one of its sub-techniques.

Second, we establish a \emph{CVE--CWE--CAPEC--ATT\&CK bridge} as shown in \figurename~\ref{fig:evidence}, which uses the CWE identifiers in $w_i$ and the available CAPEC information in $a_i$. 
A CAPEC pattern is retained when its related-weakness field shares at least one CWE identifier with the CVE. The CAPEC patterns are then projected to ATT\&CK techniques through the available CAPEC-to-ATT\&CK links.

Third, \emph{CVSS-informed retrieval} transforms the attack-vector, privileges, and user-interaction attributes in $v_i$ into natural language cues. The cue describes the vulnerability's access conditions (e.g., network-accessible and requiring no prior privileges). These descriptions are appended to the CVE query and compared with the names and descriptions of ATT\&CK techniques.

\subsubsection{Candidate Construction in Semantic Branch}

We observe that although candidates in the structural branch exploit several explicit evidence sources, they may still omit techniques whose attack semantics are expressed indirectly in the CVE description. 
Direct lexical retrieval only retrieves a technique when the CVE description shares at least one non-stopword token with the technique name or description. 
Therefore, semantically related techniques expressed using different terminology may be missed. 
In addition, CVE--CWE--CAPEC--ATT\&CK bridging also cannot recover a technique when the relevant $w_i$ or $a_i$ is unavailable. 
Furthermore, CVSS-derived cues provide supporting evidence, but they do not always identify the specific attacker behavior. To improve candidate coverage, the semantic branch uses the CVE description $d_i$ to retrieve indirectly related ATT\&CK techniques.

Therefore, we apply an AllenNLP BERT-based semantic role labeling (SRL) model~\cite{allennlp_models} to $d_i$ to extract predicate--argument structures. 
The model identifies actions expressed by the main predicates and associates them with their arguments, such as the actor, affected object, exploitation condition, instrument, and resulting effect. 
These elements are combined into compact behavioral phrases that preserve the central action and its context. For example, a description stating that a remote attacker can execute arbitrary commands through a vulnerable service may produce phrases such as "attacker executes arbitrary commands".
The extracted behavioral phrases and the ATT\&CK technique descriptions are encoded using ATT\&CK-BERT~\cite{attack_bert_huggingface}. Cosine similarity is computed between each extracted phrase and each technique description. 
Considering that a CVE may express multiple behaviors, the semantic support score of each technique is defined as the maximum phrase-level similarity across all phrases extracted from $d_i$. 
Techniques are then ranked by semantic score $z_{ij}$. 
The top 96 techniques out of 625 \A techniques that include sub-techniques are retained in the semantic candidate set $C_i^{\mathrm{sem}}$. Finally, $C_{i}^{\mathrm{struct}}$ and $C_i^{\mathrm{sem}}$ are merged to form the raw candidate set $C_{i}^{\mathrm{raw}}$.

For each candidate $c_{ij}\in C_{i}^{\mathrm{raw}}$, we construct a 19-dimensional feature vector $\phi_{ij}$ that summarizes the evidence supporting its retrieval. The features fall into four groups. 

(i) Five structural-retrieval features characterize the amount of structured evidence, including the number of retrieval sources supporting the candidate, direct lexical relevance, CWE--CAPEC association strength, parent-level lexical relevance, and token overlap between the CVE and technique text. These features measure how the CVE text matches the candidate technique and how many relevant tokens they share. 

(ii) Four semantic features capture the maximum similarity between SRL-extracted behavioral phrases and the technique description, token overlap, and the number of behavioral phrases extracted from the CVE. The \texttt{cvss\_semantic\_score} measures the semantic similarity between the technique text and a natural-language cue rendered from the CVSS attack vector, privilege, and user-interaction attributes.
They provide behavioral evidence when lexical correspondence alone is insufficient. 

(iii) Six binary provenance indicators record whether a candidate was retrieved through CWE--CAPEC bridging, direct lexical retrieval, parent-level lexical retrieval, SRL-based semantic retrieval, CVSS-informed retrieval, or CWE-based retrieval. These indicators allow the model to distinguish the presence of evidence from its numerical strength.

(iv) Four CVSS-derived binary features encode whether the vulnerability is network-accessible, requires no prior privileges, requires user interaction, or requires no user interaction. These features capture exploitation-context information.

\subsubsection{Candidate Preservation Score Model Training}
\label{subsub:score_model}

Although $C_{i}^{\mathrm{raw}}$ is constructed to achieve high candidate coverage, its broad structural and semantic retrieval strategies may introduce many weakly related or irrelevant techniques. 
Passing the entire raw candidate set to Step~2 would unnecessarily enlarge the reranking space and make the reranker more susceptible to noisy candidates. Therefore, we train a lightweight neural network model ($F_{\theta}$) as a candidate-preservation model to further reduce the Step~2 search space while retaining candidates that contain the ground-truth technique. 

For each candidate technique $c_{ij} \in C_{i}^{\mathrm{raw}}$, the $F_{\theta}$ model takes the candidate-level feature vector $\phi_{ij}$ as input. The model is a two-hidden-layer MLP with ReLU activation and dropout. Applying a sigmoid function to the MLP output $F_{\theta}({\phi}_{ij})$ yields the candidate-preservation probability $p_{ij}$. This probability estimates whether $c_{ij}$ should be retained and passed to Step~2. The features in $\phi_{ij}$ allow $F_{\theta}$ to learn which combinations of structural and semantic retrieval sources provide more reliable evidence, thereby affecting its predicted preservation probability $p_{ij}$.

The scoring model $F_{\theta}$ is trained using the training splits of KEV~\cite{kev_mappings_explorer}, ATCM~\cite{attack_to_cve_mappings}, and ENISA~\cite{enisa_vuln_report}.
As ENISA is larger than KEV and ATCM, directly training on the combined training data could cause the optimization to be dominated by ENISA. 
Therefore, we employ dataset-balanced sampling and dataset-specific loss weighting to control the contribution of each training source. 
The final checkpoint is selected based on macro-averaged validation performance across KEV and ATCM.

For each CVE record $x_i$, let $G_i$ denote its ground-truth technique set. Before training, we compare each candidate $c_{ij}$ in the raw candidate pool with the ground-truth ATT\&CK techniques annotated for the current CVE. If $c_{ij}\in G_i$, we assign it a label of 1; otherwise, we assign it a label of 0. We denote this binary training label by $y_{ij}$.

Then, we train the candidate-preservation model $F_{\theta}$ in a single supervised stage. For each raw candidate $c_{ij}$, $F_{\theta}$ takes its feature vector $\phi_{ij}$ as input and produces a candidate score $s_{ij}$, which is converted into the preservation probability $p_{ij}$ through a sigmoid function. The binary annotation-derived label $y_{ij}$ indicates whether the candidate is a ground-truth technique for the CVE record. The binary cross-entropy (BCE) optimization objective encourages high preservation probabilities for ground-truth candidates and low preservation probabilities for incorrect candidates.

To further distinguish candidates within the same raw candidate set, we add a pairwise ranking loss. For each CVE record, the pairwise loss encourages the score $s_{ij}$ of a ground-truth candidate to be higher than that of an incorrect candidate. Thus, the BCE term learns candidate-level preservation probabilities, while the pairwise term explicitly improves the relative ordering of correct and incorrect candidates within the same candidate pool. The resulting BCE--pairwise objective is used to train the final model $F_{\theta}^{\mathrm{final}}$.

\subsubsection{Final Candidate-Pool Generation}

After BCE-pairwise training, final candidate-preservation model $F_{\theta}^{\mathrm{final}}$ computes a preservation probability $p_{ij}^{\mathrm{final}}$ for each candidate $c_{ij}\in C_{i}^{\mathrm{raw}}$. For each CVE record $x_i$, candidates are ranked in descending order of $p_{ij}^{\mathrm{final}}$. 

Before dynamic selection, \S maps all 625 ATT\&CK parent techniques and sub-techniques to 201 parent-level identifiers to ensure consistency with the ground-truth annotations and avoid redundant candidates associated with the same parent technique.
For candidates associated with the same parent technique, TRACE retains the representative having the highest preservation probability. \S then ranks the parent-level candidates in descending order of their retained probabilities.
Rather than applying a fixed probability threshold, Step~1 dynamically selects final Candidate-Pool $C_{i}^{\mathrm{final}}$ according to the relative preservation probabilities within $C_{i}^{\mathrm{raw}}$. 
Specifically, the preservation probabilities of all candidates in the same raw candidate set are normalized by their sum, producing the normalized preservation probabilities $\tilde{p}_{ij}$. 
Starting from the highest-ranked candidate, Step~1 retains candidates until the cumulative normalized probability reaches the target value $\tau$ = 0.90, while limiting the pool to at most $K_{\max}$ candidates. 
If the cumulative normalized probability does not reach $\tau$ within this limit, the top $K_{\max}$ = 80 candidates are retained. 
We select $\tau$ and $K_{\max}$ using the held-out validation split by balancing ground-truth technique coverage against the number of candidates passed to Step~2. 

\subsection{Step~2: Candidate Reranking}

\begin{algorithm}[t]
\caption{Step~2: Candidate Reranking}
\label{alg:Step2}
\KwIn{
Final candidate pool $C_i^{\mathrm{final}}$;
Candidate features $\phi_{ij}$;
Step~1 preservation probabilities $p_{ij}^{\mathrm{final}}$; CVE description $d_i$; CWE information $w_i$; CVSS attributes $v_i$
}
\KwOut{Ranked candidate list $\eta_i^{\mathrm{final}}$}

$q_i \leftarrow [d_i; w_i; v_i]$\;

$\mathbf{q}_i \leftarrow \operatorname{SecRoBERTa}(q_i)$\;

\ForEach{$c_{ij}\in C_i^{\mathrm{final}}$}{

$t_{ij} \leftarrow [\operatorname{Name}(c_{ij}); \operatorname{Description}(c_{ij})]$\;

$\mathbf{c}_{ij} \leftarrow \operatorname{SecRoBERTa}(t_{ij})$\;

$\psi_{ij}^{\mathrm{text}} \leftarrow
[\mathbf{q}_i; \mathbf{c}_{ij};
|\mathbf{q}_i-\mathbf{c}_{ij}|;
\mathbf{q}_i\odot\mathbf{c}_{ij}]$\;

$u_{ij} \leftarrow h_{\omega}
\left( [\phi_{ij}; p_{ij}^{\mathrm{final}}; \psi_{ij}^{\mathrm{text}}] \right)
- \lambda_{\mathrm{freq}}\rho(c_{ij})$\;
}

$\eta_i^{\mathrm{final}} \leftarrow \operatorname{Sort}_{\downarrow}
\left( \{(c_{ij},u_{ij}) \mid c_{ij}\in C_i^{\mathrm{final}}\} \right)$\;

\Return{$\eta_i^{\mathrm{final}}$}

\end{algorithm}

Step~2 reranks the final candidate pool $C_i^{\mathrm{final}}$ produced by Step~1. Each candidate $c_{ij}\in C_i^{\mathrm{final}}$ is provided with its feature vector $\phi_{ij}$ and Step~1 preservation probability $p_{ij}^{\mathrm{final}}$.
Although Step~1 constructs a compact candidate pool while preserving high coverage, its objective is to determine whether a candidate should be retained rather than to establish the final ranking. Consequently, multiple plausible techniques supported by different structural or semantic evidence may receive similarly high preservation probabilities. However, these techniques may differ substantially in how well their attack behaviors are supported by the complete vulnerability evidence.

Furthermore, the Step~1 preservation probability $p_{ij}^{\mathrm{final}}$ alone cannot fully resolve this ambiguity because it estimates whether a candidate should be retained rather than its relative relevance among the retained candidates. Therefore, Step~2 performs reranking over $C_i^{\mathrm{final}}$ by directly modeling the semantic relationship between the CVE description and each candidate technique while incorporating the evidence accumulated in Step~1. Specifically, we train a scoring head $h_{\omega}$ on top of frozen SecRoBERTa~\cite{secroberta} representations to assign a ranking score to each retained candidate.

\begin{table*}[t]
\centering
\caption{Summary of the datasets used for training and evaluation. Source CVEs denote the number of CVEs in the original dataset. Effective CVEs denote the number remaining after cross-dataset deduplication and used in our experiments. For external test datasets, it corresponds to the actual evaluation size after removing CVEs overlapping with the Step~2 training and validation partitions. Avg.\ labels denote the average number of unique ATT\&CK technique labels per effective CVE.}
\label{tab:dataset_statistics}
\begin{tabular}{lrrrl}
\toprule
Dataset & Source CVEs & Effective CVEs & Avg.\ labels & Role \\
\midrule
KEV~\cite{kev_mappings_explorer}
& 296 & 296 & 2.69
& Step~1/2 train/val/test (178/59/59) \\

ATCM~\cite{attack_to_cve_mappings}
& 836 & 374 & 2.44
& Step~1/2 train/val/test (224/75/75) \\

ENISA~\cite{enisa_vuln_report}
& 8,072 & 8,019 & 8.03
& Step~1/2 train/val/test (4,811/1,604/1,604) \\

CVE2Attack~\cite{a15090314}
& 1,661 & 1,329 & 1.74
& External test \\

SMET~\cite{abdeen2023smet}
& 303 & 270 & 1.47
& External test \\

CGVM~\cite{liu-etal-2023-end}
& 31,298 & 27,859 & 5.48
& External test \\

\bottomrule
\end{tabular}
\end{table*}

\subsubsection{Candidate Reranker}
\label{sec:stage2_reranker}

We introduce how the final scoring head $h_{\omega}$ incorporates CVE--technique semantic relevance with Step~1 evidence to compute the final candidate score $u_{ij}$.

As illustrated in Algorithm~\ref{alg:Step2}, we first construct the CVE input $q_i$ by combining the CVE description $d_i$, CWE information $w_i$, and CVSS attributes $v_i$. 
For each retained candidate $c_{ij}$, let $t_{ij}$ denote the concatenation of its ATT\&CK technique name and description. 
SecRoBERTa separately encodes $q_i$ and $t_{ij}$ into the CVE representation $\mathbf{q}_i$ and candidate-technique representation $\mathbf{c}_{ij}$, respectively. 
We then derive the CVE--technique textual interaction feature vector $\psi_{ij}^{\mathrm{text}}$ from these two representations. 
Specifically, $\mathbf{q}_i$ and $\mathbf{c}_{ij}$ are combined through concatenation, element-wise absolute difference, and element-wise multiplication to construct $\psi_{ij}^{\mathrm{text}}$. This interaction feature captures the semantic relevance and fine-grained semantic interactions between the CVE and the candidate technique.

Finally, the Step~1 candidate feature vector $\phi_{ij}$, the Step~1 preservation probability $p_{ij}^{\mathrm{final}}$, and the CVE-technique textual interaction feature vector $\psi_{ij}^{\mathrm{text}}$ are jointly provided to the SecRoBERTa-based scoring head to compute the final candidate score:
$$
u_{ij} = h_{\omega}\!\left([\phi_{ij}; p_{ij}^{\mathrm{final}}; \psi_{ij}^{\mathrm{text}}] \right) - \lambda_{\mathrm{freq}} \rho(c_{ij}).
$$

Here, $h_{\omega}(\cdot)$ is the learned SecRoBERTa-based scoring head, which jointly models the Step~1 evidence and CVE–technique textual interactions. To reduce systematic preference toward frequently observed techniques during reranking, we further apply a frequency-based score correction. The term $\rho(c_{ij})$ denotes the training-set frequency statistic of technique $c_{ij}$, while $\lambda_{\mathrm{freq}}$ controls the strength of the corresponding penalty against systematically over-ranking frequent, generic techniques.

\subsubsection{Score Head Training}

The scoring head $h_{\omega}$ is trained to rank ground-truth ATT\&CK techniques above other candidates retained by Step~1. 
We use the same training datasets as those described in Section~\ref{subsub:score_model}.
Unlike the candidate-preservation model in Step~1, which estimates whether a candidate should remain in the candidate pool, the scoring head directly optimizes the relative ordering of the retained candidates. This distinction allows Step~1 to focus on preserving potentially relevant techniques, while Step~2 learns to place the most relevant techniques at higher ranks.

For each CVE, retained ground-truth techniques are treated as positive candidates, while the remaining retained techniques are treated as negative candidates. We construct positive--negative candidate pairs within the same candidate pool and train the scoring head using a margin-based pairwise ranking loss. The objective encourages each ground-truth technique to receive a higher score than non-ground-truth candidates by a fixed margin.
At inference time, given the Step~1 candidate pool $C_i^{\mathrm{final}}$, the final reranker computes a final score $u_{ij}$ for each retained candidate $c_{ij}$. The Rerank module then ranks all candidates in $C_i^{\mathrm{final}}$ in descending order of $u_{ij}$. The resulting ordered list, denoted by $\eta_i^{\mathrm{final}}$, constitutes the final ATT\&CK prediction.

\subsection{TRACE for Unseen CVEs: An Example}

A newly disclosed CVE may contain a vulnerability description, CVSS attributes, and CWE identifiers. 
For example, CVE-2026-73570~\cite{cve_2026_73570}, disclosed in August 2026, affects Zimbra Collaboration Suite and allows an unauthenticated attacker to execute operating system commands through SMTP requests. Its CVSS is $3.1/AV:N/AC:H/PR:N/UI:N/S:C/C:H/I:H/A:L$, indicating network-based exploitation without required privileges or user interaction. The corresponding CWE-78 identifies the underlying OS command injection weakness.

Given this information, TRACE retrieves potential ATT\&CK techniques through structural and semantic retrieval paths. 
For this CVE, CWE-78 is associated with five CAPEC patterns, but none has a corresponding ATT\&CK relationship. Meanwhile, semantic retrieval identifies additional candidates using SRL from the CVE description and natural-language representations of CVSS.

Among the Step~1 retrieved candidates, T1059.003 and T1190 receive the most support from multiple evidence sources. Specifically, T1059.003 is supported by direct lexical matching, semantic retrieval, and CWE direct lexical matching. T1190 is supported by direct lexical evidence, CWE direct lexical matching, and CVSS-based semantic matching.
Other candidates generally rely on fewer or less directly relevant evidence sources. 
For each retrieved technique, TRACE constructs an independent 19-dimensional feature vector. 
Each vector records the evidence associated with that specific candidate. For example, T1059.003 is represented by its direct lexical score of 0.7561, semantic similarity of 0.5722, and active source indicators for direct lexical retrieval, semantic retrieval, and CWE matching. 
The 19-dimensional vectors are then independently fed into the previously trained Step~1 MLP, which outputs a preservation probability for each technique candidate. 
The T1059.003 receives a preservation probability of 0.7661, the highest within the T1059 family, while T1190 receives 0.4786. Then, TRACE maps sub-techniques to their parent-level techniques and retains the highest-scoring representative within each family, along with its preservation probability and features. Thus, T1059 will proceed to Step~2. TRACE then normalizes these probabilities and selects a dynamic candidate pool. For this CVE, Step~1 retains 78 parent techniques, with T1059 and T1190 ranked first and second.

In Step~2, each retained technique is represented by its inherited Step~1 evidence features, preservation probability, and additional scoring features, forming a 23-dimensional structured feature vector. 
Meanwhile, SecRoBERTa separately encodes the CVE description, CWE information, and CVSS-derived textual representation, as well as the name and description of each parent technique. 
The resulting CVE--technique semantic interactions are combined with the Step~1 features. Meanwhile, SecRoBERTa separately encodes the CVE description, CWE identifiers, and CVSS-derived textual representation, as well as the parent technique's identifier, name, and description. 
The resulting 3,072-dimensional CVE--technique interaction vector is projected to 64 dimensions through a two-layer fusion network. This representation is concatenated with the 23-dimensional scalar features, forming an 87-dimensional input to the trained scoring head. 
The scoring head assigns a ranking score to each candidate. TRACE then sorts the candidates to produce the final ATT\&CK ranking. For CVE-2026-73570, T1059 remains ranked first, and T1190 remains ranked second.

From manual analysis, T1059 is the technique that best reflects this vulnerability. 
The description states that an unauthenticated remote attacker can supply crafted input that results in the execution of operating system commands on the affected system. 
What the attacker ultimately gains is the ability to run arbitrary commands through the system's command interpreter, which is exactly the behavior T1059 describes.

\section{EVALUATION}
\subsection{Datasets}
As shown in Table~\ref{tab:dataset_statistics}, we evaluate the framework \S on six CVE-to-ATT\&CK mapping datasets. Steps~1 and 2 are trained using KEV, ATCM, and ENISA, whereas CVE2Attack, SMET, and CGVM are used exclusively for external evaluation.

\begin{figure}[t]
\centering
\includegraphics[width=0.4\textwidth]{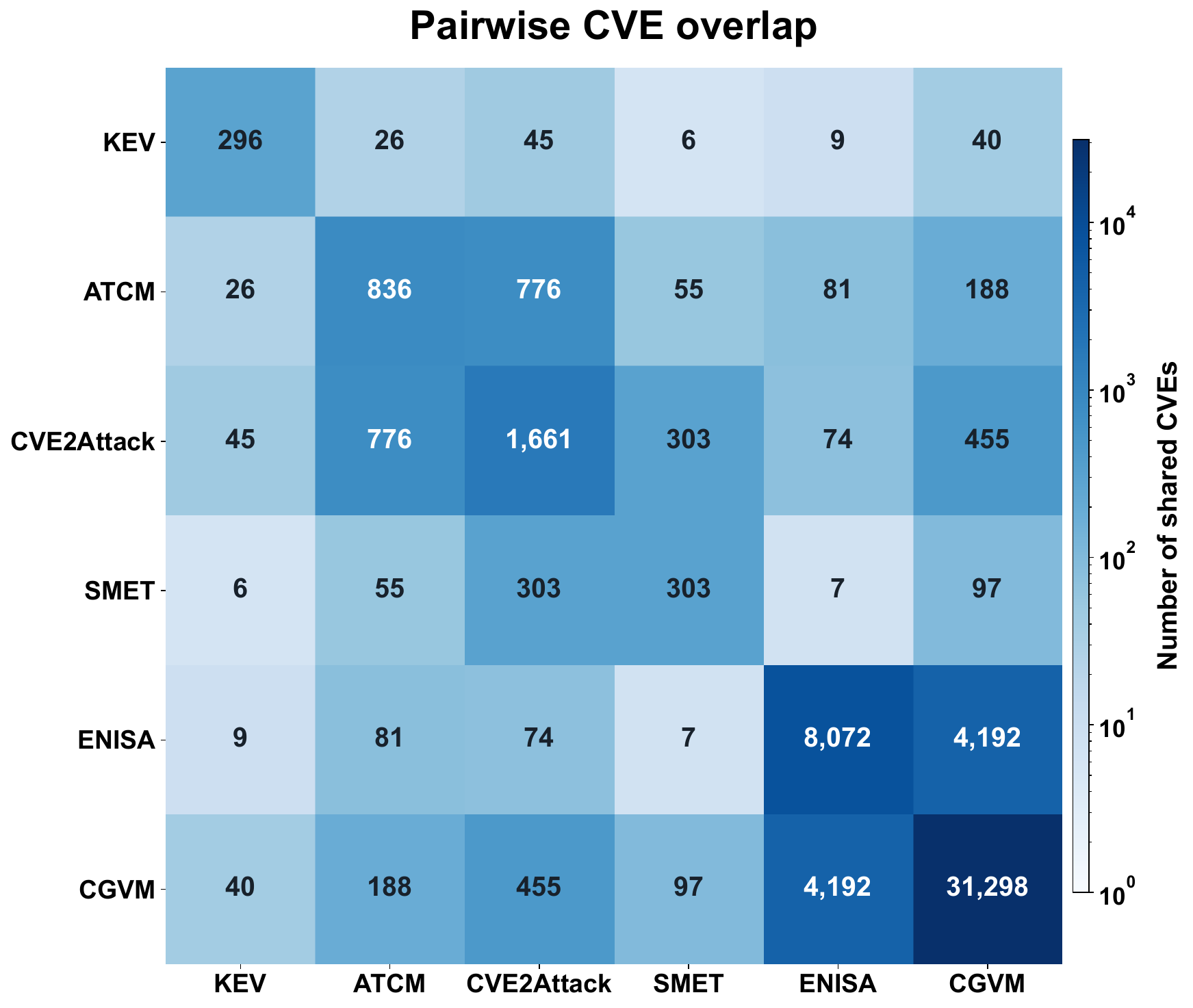}
\caption{Pairwise CVE overlap across the six datasets. Each off-diagonal cell reports the number of CVEs shared by the corresponding pair of datasets, while each diagonal cell reports the total number of unique CVEs in that dataset. Colors are shown on a logarithmic scale to accommodate the substantial differences in dataset size.}
\label{fig:overlap}
\vspace{-10pt}
\end{figure}

\subsection{Dataset Overlap Analysis}
To characterize cross-dataset relationships and control potential evaluation leakage, we analyze overlap among the six benchmarks at the CVE level. Figure~\ref{fig:overlap} shows the pairwise CVE overlap across the six datasets.

The largest pairwise overlap occurs between ENISA and CGVM, which share \(4{,}192\) CVEs. Among the shared CVEs, \(3{,}724\) have at least one common parent-level ATT\&CK technique label. CVE2Attack and ATCM also exhibit substantial overlap, sharing \(776\) CVEs, of which \(773\) have at least one common parent-level technique label. In addition, all \(303\) CVEs in SMET are also present in CVE2Attack. However, CVE-level overlap does not necessarily imply identical annotations, as different datasets may assign different ATT\&CK technique labels to the same CVE.

In contrast, KEV has relatively limited overlap with the other datasets. It shares only \(9\) CVEs with ENISA, none of which have a common parent-level technique label, and \(40\) CVEs with CGVM, of which only \(3\) share at least one label. 

These overlaps are explicitly considered in the experimental protocol. Dataset splitting and deduplication are performed at the CVE level, and CVEs appearing in the Step~2 training or validation partitions are removed from external evaluation sets. Consequently, the reported external evaluation results do not treat CVEs observed during model development as unseen test instances.

\begin{figure}[t]
\centering
\includegraphics[width=0.98\linewidth]{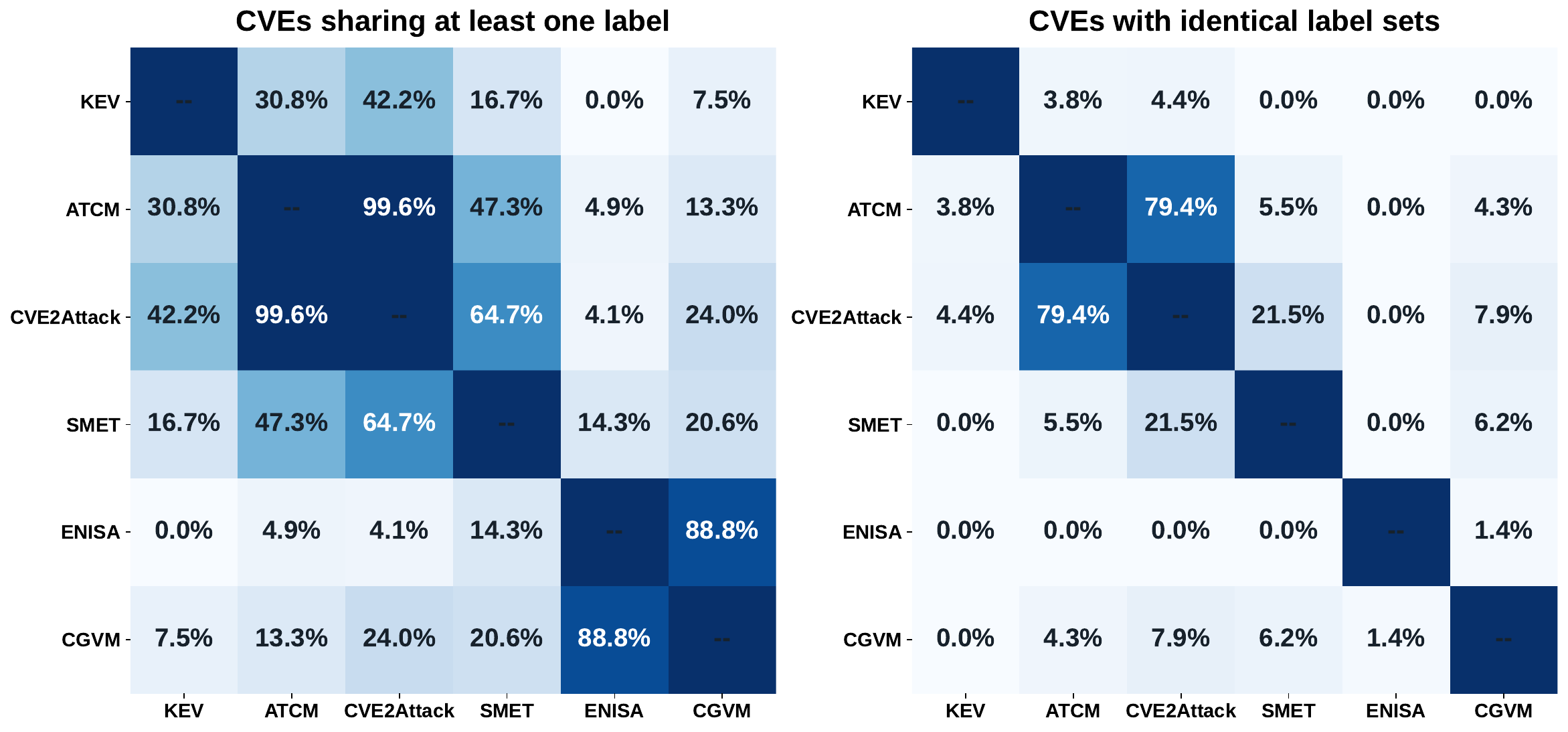}
\caption{ATT\&CK annotation consistency for CVEs shared across the six benchmark datasets. The left figure reports the proportion of shared CVEs for which each dataset pair has at least one parent-technique label in common. The right figure reports the proportion for which the complete parent-technique label are identical. The number of shared CVEs is reported separately in Figure~\ref{fig:overlap}. }
\label{fig:label_overlap}
\vspace{-10pt}
\end{figure}

As shown in Figure~\ref{fig:label_overlap}, CVE-level overlap does not necessarily imply annotation consistency. ATCM and CVE2Attack show the strongest agreement: \(99.61\%\) of their shared CVEs have at least one common parent-technique label, and \(79.38\%\) have identical label sets. In contrast, ENISA and CGVM share at least one label for \(88.84\%\) of their overlapping CVEs, but only \(1.41\%\) have identical label sets. CVE2Attack and SMET show moderate agreement, whereas KEV has relatively low agreement with several external datasets.

These differences indicate that the same CVE may receive different ATT\&CK labels across datasets. Consequently, candidates supported by similar evidence may be treated as positive examples in one dataset but as negative examples in another. Cross-dataset performance therefore reflects both model generalization and differences in annotation scope and labeling criteria.

\subsection{Evaluation Design}

We conduct experiments to answer the following questions.

\begin{itemize}
    \item \textbf{RQ1:} Can Step~1 preserve relevant techniques while maintaining a compact candidate pool?

    \item \textbf{RQ2:} Can the proposed reranker improve candidate ordering across different datasets?

    \item \textbf{RQ3:} Is candidate restriction necessary for controllable CVE to ATT\&CK mapping?

    \item \textbf{RQ4:} How does the proposed framework compare with existing baselines?
\end{itemize}

\paragraph{Evaluation metrics:}
(i) Normalized Discounted Cumulative Gain at 10 (NDCG@10)~\cite{10.1145/582415.582418} measures ranking quality within the top 10 while assigning greater importance to higher positions.
(ii) Mean Reciprocal Rank (MRR) computes the reciprocal rank of the first ground-truth technique for each CVE and reports the mean across the evaluation set.
(iii) Hit Rate at 10 (Hit@10) indicates whether at least one ground-truth technique appears in the top 10.

\subsubsection{RQ1: Candidate Pool Quality}

\begin{table}[t]
\centering
\small
\caption{Candidate-pool quality of the Step~1 with $\tau=0.90$.}
\label{tab:step1_best}
\begin{tabular}{lrrr}
\toprule
Dataset & Full Cov. & Partial Cov. & Avg.\ Cand. \\
\midrule
KEV        & 0.6780 & 1.0000 & 80.00 \\
ATCM       & 0.7067 & 0.9600 & 80.00 \\
CVE2Attack & 0.8412 & 0.9511 & 79.93 \\
SMET       & 0.9333 & 0.9778 & 79.81 \\
ENISA      & 0.4395 & 0.9994 & 79.67 \\
CGVM       & 0.7619 & 0.9705 & 79.63 \\
\bottomrule
\end{tabular}

\par\vspace{6pt}
\begin{minipage}{0.95\columnwidth}
\footnotesize
\raggedright
\textit{Note.} Full Cov.\ denotes the proportion of CVEs for which all ground-truth techniques are retained. Partial Cov.\ denotes the proportion of CVEs for which at least one ground-truth technique is retained. Avg.\ Cand.\ denotes the average number of retained candidates per CVE.
\end{minipage}
\end{table}

\begin{table*}[t]
\centering
\small
\caption{
Effect of Step~2 reranking over the same dynamically selected candidate pools.
}
\label{tab:step1_step2_comparison}
\resizebox{\linewidth}{!}{
\begin{tabular}{lccccc|ccccc|ccccc}
\toprule
& \multicolumn{5}{c|}{Step~1 Candidate Ordering}
& \multicolumn{5}{c|}{Step~2 Reranking}
& \multicolumn{5}{c}{Gain} \\
\cmidrule(lr){2-6}
\cmidrule(lr){7-11}
\cmidrule(l){12-16}
Dataset
& NDCG@10 & MRR & Hit@1 & Hit@5 & Hit@10
& NDCG@10 & MRR & Hit@1 & Hit@5 & Hit@10
& $\Delta$NDCG & $\Delta$MRR & $\Delta$Hit@1
& $\Delta$Hit@5 & $\Delta$Hit@10 \\
\midrule

KEV
& 0.3236 & 0.4196 & 0.2712 & 0.5932 & 0.7627
& \textbf{0.6042} & \textbf{0.7404} & \textbf{0.6271}
& \textbf{0.8983} & \textbf{0.9661}
& +0.2806 & +0.3207 & +0.3559 & +0.3051 & +0.2034 \\

ATCM
& 0.2912 & 0.3476 & 0.2133 & 0.5467 & 0.6533
& \textbf{0.5449} & \textbf{0.5816} & \textbf{0.4267}
& \textbf{0.8133} & \textbf{0.9200}
& +0.2537 & +0.2340 & +0.2133 & +0.2667 & +0.2667 \\

CVE2Attack
& 0.3631 & 0.3902 & 0.2626 & 0.5320 & 0.6697
& \textbf{0.4268} & \textbf{0.4319} & \textbf{0.2822}
& \textbf{0.6110} & \textbf{0.7457}
& +0.0637 & +0.0417 & +0.0196 & +0.0790 & +0.0760 \\

SMET
& \textbf{0.4618} & \textbf{0.4414} & \textbf{0.2704}
& \textbf{0.6741} & \textbf{0.8074}
& 0.4399 & 0.4127 & 0.2407 & 0.6222 & 0.7741
& $-0.0220$ & $-0.0286$ & $-0.0296$ & $-0.0519$ & $-0.0333$ \\

ENISA
& 0.2183 & 0.2282 & 0.0704 & 0.4370 & 0.5636
& \textbf{0.3097} & \textbf{0.3395} & \textbf{0.1234}
& \textbf{0.6340} & \textbf{0.8872}
& +0.0915 & +0.1113 & +0.0530 & +0.1970 & +0.3236 \\

CGVM
& 0.4113 & 0.3882 & 0.1965 & 0.6981 & 0.7995
& \textbf{0.4720} & \textbf{0.5019} & \textbf{0.3194}
& \textbf{0.7455} & \textbf{0.8495}
& +0.0607 & +0.1138 & +0.1229 & +0.0475 & +0.0500 \\

\bottomrule
\end{tabular}
}
\end{table*}

Table~\ref{tab:step1_best} reports the quality of the candidate pools produced by Step~1 on six datasets. With $\tau=0.90$, $K_{\min}=8$, and $K_{\max}=80$, Step~1 reduces the search space of 201 parent-level ATT\&CK techniques by approximately 60\%. Despite this reduction, Step~1 retains at least one ground-truth technique for 95.11\%--100\% of CVEs across the six datasets. These results show that Step~1 constructs a bounded candidate pool while maintaining high partial coverage.

In ENISA and CGVM, full-coverage rates are 43.95\% and 76.19\%, respectively, despite partial-coverage rates of 99.94\% and 97.05\%. This gap is primarily due to their denser multi-label annotations, averaging 8.03 and 5.48 ground-truth techniques per CVE, respectively. Since full coverage requires retaining every ground-truth technique, missing even one label results in a full-coverage failure, while partial coverage only requires retaining at least one. Consequently, high partial coverage can coexist with lower full coverage on densely annotated datasets.


\subsubsection{RQ2: Reranking Performance across Datasets}
\label{subsec:reranking_results}

Table~\ref{tab:step1_step2_comparison} compares the Step~1 candidate ordering with Step~2 reranking over exactly the same dynamically selected candidate pools. Thus, all observed differences are attributable to reranking rather than to the retrieval of additional techniques. Step~2 improves all five ranking metrics on five of the six datasets. The largest gains occur on KEV and ATCM: NDCG@10 increases from 0.3236 to 0.6042 on KEV and from 0.2912 to 0.5449 on ATCM. The improvement is also substantial on ENISA, where Hit@10 increases from 0.5636 to 0.8872.

SMET is the only exception, with NDCG@10 decreasing from 0.4618 to 0.4399 and Hit@10 decreasing from 0.8074 to 0.7741. This behavior reflects a benchmark-specific semantic mismatch. SMET follows a semantic attack-vector mapping procedure: CVE descriptions are decomposed into attack vectors and mapped to ATT\&CK techniques through semantic correspondence with ATT\&CK attack vectors. In contrast, Step~2 uses SecRoBERTa-based CVE--technique interaction representations. This difference in semantic representation can alter the ordering of related techniques and reduce agreement with SMET's labeling criteria, even when the candidate pool contains relevant techniques. Accordingly, the SMET decrease should be interpreted as sensitivity to annotation semantics rather than a failure of candidate-pool preservation.

\subsubsection{RQ3: Effectiveness of Candidate-Pool Restriction}

\begin{table*}[t]
\centering
\small
\caption{Effect of candidate-pool construction on LLM-based CVE-to-ATT\&CK mapping performance.}
\label{tab:candidate_pool_llm}
\resizebox{\linewidth}{!}{
\begin{tabular}{lccccc|ccccc|ccccc}
\toprule
& \multicolumn{5}{c|}{Without Candidate Pool}
& \multicolumn{5}{c|}{With Candidate Pool}
& \multicolumn{5}{c}{Gain} \\
\cmidrule(lr){2-6}
\cmidrule(lr){7-11}
\cmidrule(l){12-16}

Model

& NDCG@10 & MRR & Hit@1 & Hit@5 & Hit@10
& NDCG@10 & MRR & Hit@1 & Hit@5 & Hit@10
& $\Delta$NDCG & $\Delta$MRR & $\Delta$Hit@1
& $\Delta$Hit@5 & $\Delta$Hit@10 \\

\midrule
GPT-4o-mini

& 0.3553 & 0.5236 & 0.4291 & 0.6318 & 0.6959
& \textbf{0.5172} & \textbf{0.6767} & \textbf{0.5473}
& \textbf{0.8615} & \textbf{0.9155}
& +0.1619 & +0.1531 & +0.1182 & +0.2297 & +0.2196 \\

GPT-5.4-mini
& 0.5241 & 0.6665 & 0.5507 & 0.8345 & 0.9155
& \textbf{0.5329} & \textbf{0.6904} & \textbf{0.5574}
& \textbf{0.8682} & \textbf{0.9324}
& +0.0088 & +0.0239 & +0.0068 & +0.0338 & +0.0169 \\
\bottomrule
\end{tabular}
}
\end{table*}


Table~\ref{tab:candidate_pool_llm} reports the results on the KEV dataset. Restricting the mapping space to the Step~1 candidate pool improves all ranking metrics for both LLMs. For GPT-4o-mini, NDCG@10 increases from 0.3553 to 0.5172 and MRR increases from 0.5236 to 0.6767. Hit@10 also improves from 0.6959 to 0.9155. GPT-5.4-mini also improves consistently across all five metrics. Although the gains are smaller than those observed for GPT-4o-mini, its Hit@10 increases from 0.9155 to 0.9324. These results show that the benefit of Step~1 is not limited to our Step~2 reranker. Restricting the search space to the retained candidates can also improve the ranking performance of general-purpose LLMs by reducing interference from irrelevant ATT\&CK techniques while preserving relevant candidates.

\subsubsection{RQ4: Comparison with Baselines}





\begin{figure}[!t]
\centering
\includegraphics[width=0.45\textwidth]{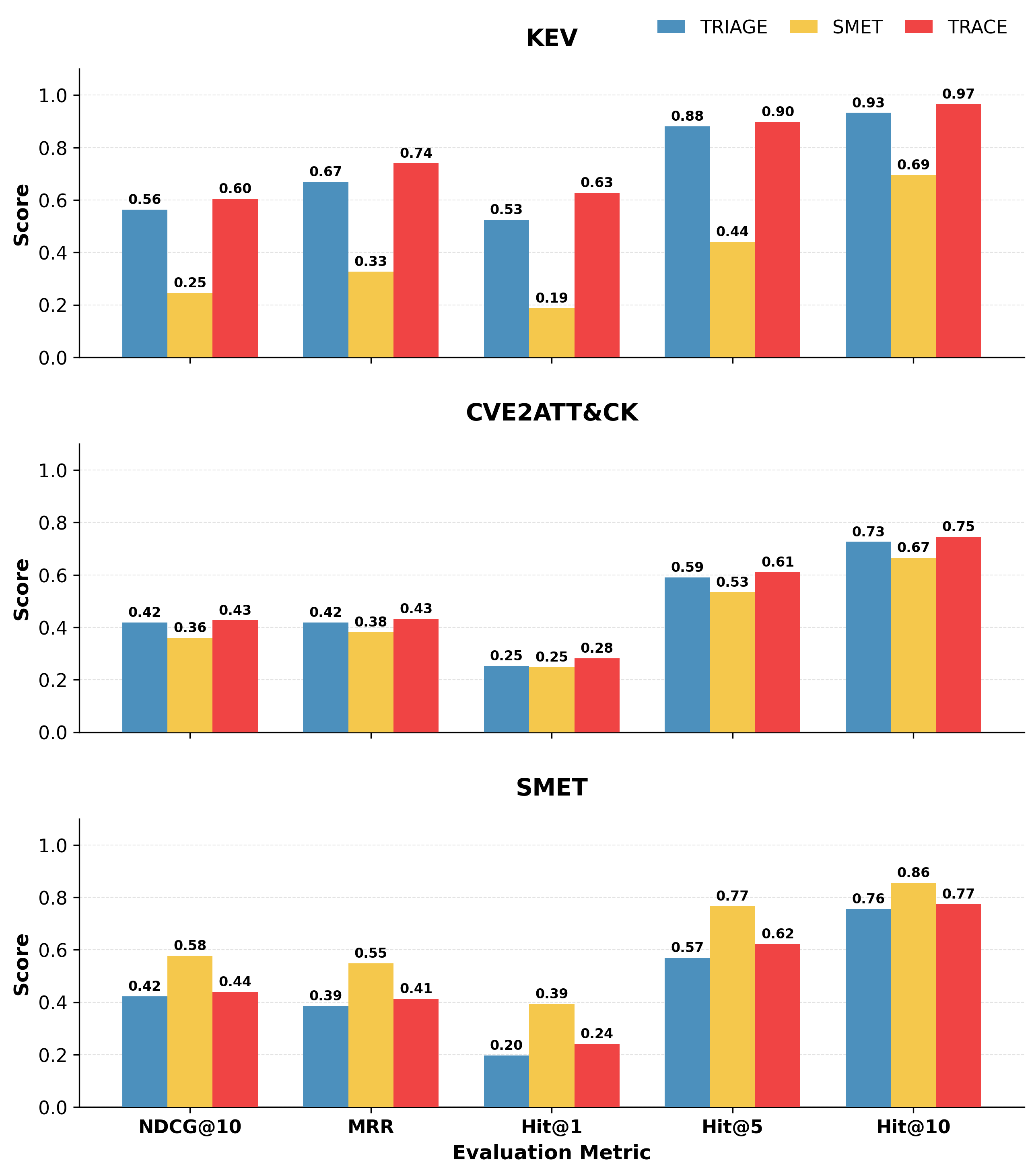}
\caption{Comparison of TRACE with the TRIAGE and SMET baselines.
}
\label{baseline}
\vspace{-10pt}
\end{figure}

Figure~\ref{baseline} compares our framework \S with TRIAGE~\cite{host2025systematicapproachpredictimpact} and SMET~\cite{abdeen2023smet}. TRACE achieves the best performance across all five metrics on both KEV and CVE2ATT\&CK. On KEV, TRACE reaches an NDCG@10 of 0.6042 and an MRR of 0.7404, compared with 0.5627 and 0.6689 for TRIAGE. On CVE2ATT\&CK, TRACE also achieves the highest ranking scores, including an NDCG@10 of 0.4268 and a Hit@10 of 0.7457, demonstrating its effectiveness across datasets with different annotation distributions.

On the SMET dataset, the SMET baseline achieves the strongest performance, with an NDCG@10 of 0.5777 and a Hit@10 of 0.8556, compared with 0.4399 and 0.7741 for TRACE. This difference may reflect the benchmark-specific semantic alignment between the SMET method and its annotation procedure, whereas TRACE integrates heterogeneous structural and semantic evidence rather than relying exclusively on semantic attack-vector correspondence.

\subsection{Ablation Study}

We conduct ablation studies to isolate the contributions of the two-step design. For Step~1, we separately remove the structural, semantic retrieval branches, and direct lexical retrieval to assess their roles in retaining ground-truth techniques within a bounded candidate pool. For Step~2, we separately remove the SecRoBERTa-based textual interaction representation and the Step~1 candidate features to evaluate their contributions to candidate reranking. We further evaluate robustness to missing evidence at inference time by removing individual evidence sources while keeping both model checkpoints frozen. Finally, we examine generalization to long-tail techniques.

\subsubsection{Ablation Study of Step~1}

\begin{table}[t]
\centering
\caption{
Ablation study of Step~1 candidate-pool construction under the
current CVE-level merged retrieval setting.
}
\label{tab:step1_ablation}
\resizebox{\linewidth}{!}{
\begin{tabular}{llrrrr}
\toprule
Dataset & Method & Label Cov. & Full Cov. & Partial Cov. & Avg. Pool \\
\midrule

KEV & Full TRACE
& \textbf{0.8387} & \textbf{0.6780}
& \textbf{1.0000} & 80.00 \\
KEV & $-$ Semantic retrieval
& 0.6516 & 0.3729 & 0.9661 & 80.00 \\
KEV & $-$ Structural retrieval
& 0.6129 & 0.3390 & 0.8305 & 35.53 \\
KEV & $-$ Direct retrieval
& 0.6194 & 0.3390 & 0.8475 & 35.81 \\
\midrule

ATCM & Full TRACE
& 0.8208 & \textbf{0.7067}
& \textbf{0.9600} & 80.00 \\
ATCM & $-$ Semantic retrieval
& \textbf{0.8266} & 0.6667 & 0.9467 & 80.00 \\
ATCM & $-$ Structural retrieval
& 0.4740 & 0.3067 & 0.6800 & 29.79 \\
ATCM & $-$ Direct retrieval
& 0.5029 & 0.2933 & 0.7600 & 30.88 \\
\midrule

CVE2Attack & Full TRACE
& \textbf{0.8988} & \textbf{0.8412} & \textbf{0.9511} & 79.93 \\
CVE2Attack & $-$ Semantic retrieval
& 0.8508 & 0.7833 & 0.9202 & 79.89 \\
CVE2Attack & $-$ Structural retrieval
& 0.5045 & 0.3702 & 0.5989 & 29.34 \\
CVE2Attack & $-$ Direct retrieval
& 0.5750 & 0.4447 & 0.6862 & 29.91 \\
\midrule

SMET & Full TRACE
& \textbf{0.9545} & \textbf{0.9333} & \textbf{0.9778} & 79.81 \\
SMET & $-$ Semantic retrieval
& 0.8712 & 0.8222 & 0.9185 & 79.61 \\
SMET & $-$ Structural retrieval
& 0.6540 & 0.5704 & 0.7111 & 32.09 \\
SMET & $-$ Direct retrieval
& 0.7298 & 0.6444 & 0.8074 & 32.66 \\
\midrule

ENISA & Full TRACE
& \textbf{0.5791} & \textbf{0.4395} & \textbf{0.9994} & 79.67 \\
ENISA & $-$ Semantic retrieval
& 0.5607 & 0.4133 & 0.9956 & 79.18 \\
ENISA & $-$ Structural retrieval
& 0.2055 & 0.1640 & 0.6590 & 30.22 \\
ENISA & $-$ Direct retrieval
& 0.3894 & 0.3454 & 0.8847 & 32.96 \\
\midrule

CGVM & Full TRACE
& \textbf{0.7803} & \textbf{0.7619} & \textbf{0.9705} & 79.63 \\
CGVM & $-$ Semantic retrieval
& 0.7476 & 0.6818 & 0.9465 & 79.20 \\
CGVM & $-$ Structural retrieval
& 0.2446 & 0.2508 & 0.6189 & 27.49 \\
CGVM & $-$ Direct retrieval
& 0.6264 & 0.6305 & 0.8908 & 30.96 \\

\bottomrule
\end{tabular}
}
\par\vspace{6pt}
\footnotesize
\raggedright
\textit{Note.}
Label Cov.\ is the proportion of individual ground-truth parent-technique labels retained in the Step~1 dynamic candidate pool.
\end{table}

Table~\ref{tab:step1_ablation} evaluates the contributions of the structural and semantic retrieval branches to candidate-pool construction. Removing the structural branch consistently causes the largest coverage loss. For example, on CVE2Attack, Label Coverage decreases from 0.8988 to 0.5045, while Partial Coverage decreases from 0.9511 to 0.5989, demonstrating that structural evidence provides the principal source of candidate recall.

The effect of the semantic branch is dataset dependent. It substantially improves coverage on KEV and SMET and recovers candidates that are absent from the structural paths. Overall, the structural branch supplies the majority of candidate recall, whereas the semantic branch provides complementary but distribution-dependent evidence.

\subsubsection{Ablation Study of Step~2}

\begin{table}[t]
\centering
\caption{Ablation study of the Step~2 reranker under the CVE-level merged candidate-pool setting.}
\label{tab:step2_component_ablation}
\resizebox{\linewidth}{!}{
\begin{tabular}{llrrrrr}
\toprule
Dataset & Method
& NDCG@10 & MRR & Hit@1 & Hit@5 & Hit@10 \\
\midrule

\multirow{3}{*}{KEV}
& Full model
& \textbf{0.6042} & \textbf{0.7404} & \textbf{0.6271}
& \textbf{0.8983} & \textbf{0.9661} \\
& w/o SecRoBERTa
& 0.3433 & 0.4457 & 0.3051 & 0.5254 & 0.7797 \\
& w/o Step~1 evidence
& 0.5932 & 0.6956 & 0.5593 & 0.8814 & 0.9153 \\
\midrule

\multirow{3}{*}{ATCM}
& Full model
& \textbf{0.5449} & \textbf{0.5816} & \textbf{0.4267}
& \textbf{0.8133} & \textbf{0.9200} \\
& w/o SecRoBERTa
& 0.3002 & 0.3418 & 0.1867 & 0.5467 & 0.6800 \\
& w/o Step~1 evidence
& 0.4846 & 0.5199 & 0.3733 & 0.7067 & 0.8000 \\
\midrule

\multirow{3}{*}{CVE2Attack}
& Full model
& \textbf{0.4268} & \textbf{0.4319} & \textbf{0.2822}
& \textbf{0.6110} & \textbf{0.7457} \\
& w/o SecRoBERTa
& 0.3198 & 0.3416 & 0.2160 & 0.4808 & 0.6275 \\
& w/o Step~1 evidence
& 0.3556 & 0.3727 & 0.2378 & 0.5305 & 0.6479 \\
\midrule

\multirow{3}{*}{SMET}
& Full model
& \textbf{0.4399} & \textbf{0.4127} & \textbf{0.2407}
& \textbf{0.6222} & \textbf{0.7741} \\
& w/o SecRoBERTa
& 0.3661 & 0.3512 & 0.2037 & 0.5111 & 0.6815 \\
& w/o Step~1 evidence
& 0.3377 & 0.3206 & 0.1593 & 0.5000 & 0.6444 \\
\midrule

\multirow{3}{*}{ENISA}
& Full model
& 0.3097 & 0.3395 & 0.1234 & \textbf{0.6340} & \textbf{0.8872} \\
& w/o SecRoBERTa interaction
& 0.2815 & 0.3434 & 0.1870 & 0.5237 & 0.6652 \\
& w/o all Step~1 evidence
& \textbf{0.3526} & \textbf{0.4047} & \textbf{0.2712} & 0.6165 & 0.8615 \\

\midrule

\multirow{3}{*}{CGVM}
& Full model
& \textbf{0.4720} & \textbf{0.5019} & \textbf{0.3194}
& \textbf{0.7455} & \textbf{0.8495} \\
& w/o SecRoBERTa
& 0.2758 & 0.2689 & 0.1197 & 0.4438 & 0.6301 \\
& w/o Step~1 evidence
& 0.3988 & 0.4569 & 0.3117 & 0.6306 & 0.7508 \\

\bottomrule
\end{tabular}
}
\end{table}

Table~\ref{tab:step2_component_ablation} evaluates the contributions of CVE--technique textual interactions and Step~1 evidence to the Step~2 reranker. Removing the SecRoBERTa interaction representation decreases NDCG@10, Hit@5, and Hit@10 across all six datasets. The largest NDCG@10 decreases occur on KEV and ATCM, from 0.6042 to 0.3433 and from 0.5449 to 0.3002, respectively. On ENISA, removing textual interactions increases MRR from 0.3395 to 0.3434 and Hit@1 from 0.1234 to 0.1870, but reduces Hit@10 from 0.8872 to 0.6652. These results demonstrate that textual interactions are essential for maintaining top-$k$ ranking performance, although their effects on the highest-ranked predictions can vary across datasets.

Removing all Step~1 evidence, including the candidate feature vector and preservation probability, decreases Hit@5 and Hit@10 across all six datasets. On SMET, NDCG@10 decreases from 0.4399 to 0.3377 and Hit@10 from 0.7741 to 0.6444. Similar reductions occur on ATCM and CVE2Attack, indicating that the retained evidence provides additional information for distinguishing relevant techniques beyond textual similarity. However, ENISA exhibits a different pattern: removing Step~1 evidence improves NDCG@10 from 0.3097 to 0.3526, MRR from 0.3395 to 0.4047, and Hit@1 from 0.1234 to 0.2712, while reducing Hit@10 from 0.8872 to 0.8615. This suggests that Step~1 evidence improves the inclusion of relevant techniques within the top-ranked results but may introduce less effective ordering signals for the highest-ranked predictions on ENISA.

Overall, both information sources consistently improve Hit@5 and Hit@10 across all six datasets, demonstrating their complementary contributions to reranking. SecRoBERTa interactions provide the stronger and more consistent contribution to NDCG@10, whereas Step~1 evidence further improves top-$k$ coverage. The mixed effects on MRR and Hit@1, particularly on ENISA, indicate that improving the inclusion of relevant techniques within the top-$k$ results does not necessarily improve their ordering at the highest ranks.

\subsubsection{Performance on Missing Evidence}

\begin{table*}[t]
\centering
\scriptsize
\caption{
Performance of TRACE under missing evidence at inference time. The Step~1 and Step~2 checkpoints remain frozen for all configurations. Macro Avg.\ denotes the average over the six datasets.
}
\label{tab:missing_evidence}
\setlength{\tabcolsep}{3.5pt}
\resizebox{0.88\linewidth}{!}{
\begin{tabular}{llrrrrrrrr}
\toprule
Dataset & Evidence setting
& Label Cov.
& Full Cov.
& Partial Cov.
& NDCG@10
& MRR
& Hit@1
& Hit@5
& Hit@10 \\
\midrule

\multirow{6}{*}{KEV}
& Full evidence
& 0.8387 & 0.6780 & 1.0000
& 0.6042 & 0.7404 & 0.6271 & 0.8983 & 0.9661 \\
& w/o CWE
& 0.8387 & 0.6780 & 1.0000
& 0.6008 & 0.7326 & 0.6102 & 0.8983 & 0.9831 \\
& w/o CVSS
& 0.6968 & 0.4068 & 0.9322
& 0.4979 & 0.6433 & 0.5254 & 0.8136 & 0.8475 \\
& w/o CAPEC
& 0.8387 & 0.6780 & 1.0000
& 0.6042 & 0.7408 & 0.6271 & 0.8983 & 0.9661 \\
& w/o SRL
& 0.6968 & 0.4746 & 0.9831
& 0.5448 & 0.6685 & 0.5424 & 0.8814 & 0.9322 \\
& w/o CWE and CVSS
& 0.6903 & 0.3729 & 0.9492
& 0.4920 & 0.6548 & 0.5424 & 0.8305 & 0.8644 \\

\midrule

\multirow{6}{*}{ATCM}
& Full evidence
& 0.8208 & 0.7067 & 0.9600
& 0.5449 & 0.5816 & 0.4267 & 0.8133 & 0.9200 \\
& w/o CWE
& 0.8000 & 0.6933 & 0.9600
& 0.5430 & 0.5746 & 0.4133 & 0.8133 & 0.9200 \\
& w/o CVSS
& 0.7771 & 0.6133 & 0.9467
& 0.4762 & 0.5263 & 0.3467 & 0.7600 & 0.8267 \\
& w/o CAPEC
& 0.8114 & 0.7067 & 0.9600
& 0.5413 & 0.5756 & 0.4133 & 0.8267 & 0.9333 \\
& w/o SRL
& 0.7771 & 0.6267 & 0.9467
& 0.4725 & 0.5154 & 0.3467 & 0.7600 & 0.8933 \\
& w/o CWE and CVSS
& 0.7771 & 0.6133 & 0.9467
& 0.4788 & 0.5216 & 0.3467 & 0.7333 & 0.8400 \\

\midrule

\multirow{6}{*}{CVE2Attack}
& Full evidence
& 0.8988 & 0.8412 & 0.9511
& 0.4268 & 0.4319 & 0.2822 & 0.6110 & 0.7457 \\
& w/o CWE
& 0.8902 & 0.8284 & 0.9436
& 0.4295 & 0.4349 & 0.2799 & 0.6238 & 0.7517 \\
& w/o CVSS
& 0.8115 & 0.7171 & 0.9014
& 0.3872 & 0.4067 & 0.2656 & 0.5854 & 0.7035 \\
& w/o CAPEC
& 0.8980 & 0.8397 & 0.9488
& 0.4263 & 0.4302 & 0.2777 & 0.6147 & 0.7555 \\
& w/o SRL
& 0.8236 & 0.7449 & 0.9097
& 0.3645 & 0.3692 & 0.2212 & 0.5433 & 0.6862 \\
& w/o CWE and CVSS
& 0.7817 & 0.6734 & 0.8789
& 0.3611 & 0.3925 & 0.2573 & 0.5621 & 0.6764 \\

\midrule

\multirow{6}{*}{SMET}
& Full evidence
& 0.9545 & 0.9333 & 0.9778
& 0.4399 & 0.4127 & 0.2407 & 0.6222 & 0.7741 \\
& w/o CWE
& 0.9571 & 0.9370 & 0.9815
& 0.4715 & 0.4498 & 0.2889 & 0.6296 & 0.7815 \\
& w/o CVSS
& 0.9040 & 0.8704 & 0.9556
& 0.3701 & 0.3474 & 0.1704 & 0.5481 & 0.7000 \\
& w/o CAPEC
& 0.9545 & 0.9333 & 0.9778
& 0.4248 & 0.3944 & 0.2185 & 0.6111 & 0.7704 \\
& w/o SRL
& 0.8965 & 0.8519 & 0.9481
& 0.3438 & 0.3154 & 0.1444 & 0.5037 & 0.6889 \\
& w/o CWE and CVSS
& 0.8939 & 0.8556 & 0.9519
& 0.3728 & 0.3689 & 0.2111 & 0.5481 & 0.6963 \\

\midrule

\multirow{6}{*}{ENISA}
& Full evidence
& 0.5791 & 0.4395 & 0.9994
& 0.3097 & 0.3395 & 0.1234 & 0.6340 & 0.8872 \\
& w/o CWE
& 0.4699 & 0.4034 & 0.9963
& 0.1718 & 0.2186 & 0.0249 & 0.4688 & 0.8198 \\
& w/o CVSS
& 0.5974 & 0.4289 & 0.9988
& 0.2851 & 0.2954 & 0.0867 & 0.5848 & 0.8435 \\
& w/o CAPEC
& 0.4680 & 0.4046 & 0.9975
& 0.1804 & 0.2200 & 0.0299 & 0.4638 & 0.8229 \\
& w/o SRL
& 0.6101 & 0.4408 & 0.9994
& 0.3524 & 0.3810 & 0.1559 & 0.7001 & 0.9021 \\
& w/o CWE and CVSS
& 0.5158 & 0.3884 & 0.9975
& 0.1483 & 0.1809 & 0.0193 & 0.3510 & 0.7132 \\

\midrule

\multirow{6}{*}{CGVM}
& Full evidence
& 0.7803 & 0.7619 & 0.9705
& 0.4720 & 0.5019 & 0.3194 & 0.7455 & 0.8495 \\
& w/o CWE
& 0.6131 & 0.5759 & 0.9477
& 0.2730 & 0.2991 & 0.1128 & 0.5593 & 0.7747 \\
& w/o CVSS
& 0.7943 & 0.7516 & 0.9635
& 0.4589 & 0.4904 & 0.3173 & 0.7224 & 0.8319 \\
& w/o CAPEC
& 0.6124 & 0.5842 & 0.9519
& 0.2832 & 0.3145 & 0.1385 & 0.5553 & 0.7753 \\
& w/o SRL
& 0.7890 & 0.7510 & 0.9592
& 0.5405 & 0.5856 & 0.4219 & 0.7957 & 0.8649 \\
& w/o CWE and CVSS
& 0.6372 & 0.5715 & 0.9446
& 0.2729 & 0.3059 & 0.1373 & 0.5326 & 0.7515 \\

\midrule

\multirow{6}{*}{Macro Avg.}
& Full evidence
& \textbf{0.8121} & \textbf{0.7268} & \textbf{0.9765}
& \textbf{0.4663} & \textbf{0.5013} & \textbf{0.3366}
& \textbf{0.7207} & \textbf{0.8571} \\
& w/o CWE
& 0.7615 & 0.6860 & 0.9715
& 0.4149 & 0.4516 & 0.2883 & 0.6655 & 0.8385 \\
& w/o CVSS
& 0.7635 & 0.6313 & 0.9497
& 0.4126 & 0.4516 & 0.2853 & 0.6690 & 0.7922 \\
& w/o CAPEC
& 0.7638 & 0.6911 & 0.9727
& 0.4100 & 0.4459 & 0.2842 & 0.6617 & 0.8373 \\
& w/o SRL
& 0.7655 & 0.6483 & 0.9577
& 0.4364 & 0.4725 & 0.3054 & 0.6974 & 0.8279 \\
& w/o CWE and CVSS
& 0.7160 & 0.5792 & 0.9448
& 0.3543 & 0.4041 & 0.2524 & 0.5929 & 0.7570 \\

\bottomrule
\end{tabular}
}
\end{table*}

Table~\ref{tab:missing_evidence} shows that missing evidence affects both candidate coverage and ranking performance. Notably, Partial Coverage decreases by only 3.17 percentage points, compared with a 14.76-point decrease in Full Coverage, indicating that missing evidence affects complete label-set recovery more strongly than the preservation of at least one relevant technique. When both CWE and CVSS are unavailable, macro-average Label Coverage decreases from 0.8121 to 0.7160, Full Coverage from 0.7268 to 0.5792, and Partial Coverage from 0.9765 to 0.9448, while Hit@10 decreases from 0.8571 to 0.7570. These results indicate that the remaining evidence sources can still recover at least one relevant technique for most CVEs, but recovering complete label sets and ranking relevant techniques highly become more challenging. The effects also differ by source: removing CVSS reduces macro-average Hit@10 from 0.8571 to 0.7922, whereas removing CWE or CAPEC particularly affects ENISA and CGVM. Overall, TRACE maintains relatively strong performance under missing evidence, preserving high partial coverage and retaining useful ranking capability despite the reduced information.

\subsubsection{Performance on Long-Tail Techniques}

\begin{table}[t]
\centering
\caption{Performance across technique-frequency groups on external evaluation datasets.}
\label{tab:train_frequency_long_tail_results}
\resizebox{\linewidth}{!}{
\begin{tabular}{llrrrrrr}
\toprule
Dataset & Group & \#Tech. & \#Gold Cases & Partial Cov. & Hit@10 & NDCG@10 & MRR \\
\midrule

\multirow{4}{*}{CVE2Attack}
& Head   & 12 & 867 & 0.9423 & 0.8316 & 0.5294 & 0.4938 \\
& Medium & 5  & 419 & 0.8783 & 0.7661 & 0.4399 & 0.3574 \\
& Tail   & 12 & 551 & 0.9238 & 0.2632 & 0.1071 & 0.0955 \\
& Unseen & 2  & 40  & 0.7000 & 0.0000 & 0.0000 & 0.0157 \\
\midrule

\multirow{4}{*}{SMET}
& Head   & 15 & 148 & 0.9527 & 0.7838 & 0.4749 & 0.4041 \\
& Medium & 4  & 105 & 0.9524 & 0.8381 & 0.5433 & 0.4607 \\
& Tail   & 16 & 107 & 0.9813 & 0.4206 & 0.1725 & 0.1294 \\
& Unseen & 5  & 6   & 1.0000 & 0.0000 & 0.0000 & 0.0269 \\
\midrule

\multirow{4}{*}{CGVM}
& Head   & 33 & 23,638 & 0.9929 & 0.9118 & 0.5128 & 0.5346 \\
& Medium & 3  & 2,340  & 0.9688 & 0.9274 & 0.6237 & 0.5770 \\
& Tail   & 18 & 11,729 & 0.8643 & 0.2297 & 0.0718 & 0.0724 \\
& Unseen & 14 & 2,039  & 0.8185 & 0.0113 & 0.0035 & 0.0219 \\
\bottomrule
\end{tabular}
}
\par\vspace{6pt}
\begin{minipage}{0.98\linewidth}
\footnotesize
\raggedright
\textit{Note.} \#Gold Cases denotes the number of CVEs containing at least one ground truth technique in each frequency group. A CVE may belong to multiple groups. Partial Cov.\ and Hit@10 measure whether at least one group-specific ground truth technique is retained in the Step~1 pool and final top~10, respectively. NDCG@10 and MRR are computed using only group-specific ground truth techniques.
\end{minipage}
\end{table}

To evaluate cross-dataset generalization to low-frequency and unseen techniques, we calculate each technique's frequency using the ground-truth labels in the pooled KEV, ATCM, and ENISA training partitions. A technique is counted at most once for each training CVE whose ground-truth set contains it. We classify techniques with more than 50 training occurrences as \emph{Head}, those with 20--50 occurrences as \emph{Medium}, those with 1--19 occurrences as \emph{Tail}, and those with no training occurrences as \emph{Unseen}. 

Table~\ref{tab:train_frequency_long_tail_results} reports the performance of these four frequency groups on the external CVE2Attack, SMET, and CGVM datasets. Step~1 retains substantial coverage of Tail and Unseen techniques across all three external datasets. For Tail techniques, Partial Cov.\ reaches 0.9238, 0.9813, and 0.8643 on CVE2Attack, SMET, and CGVM, respectively. For Unseen techniques, it remains 0.7000, 1.0000, and 0.8185. However, ranking performance degrades markedly for rare techniques. These results show that Step~1 can often retain low-frequency and unseen techniques in the candidate pool, while accurately prioritizing them remains challenging.

\subsection{Hyperparameter Study}

\subsubsection{Sensitivity to $\tau$}
The score-mass threshold $\tau$ is used during Step~1 inference to determine the number of candidates retained for each CVE. After candidates are sorted by their preservation scores, the scores are normalized to sum to one. Step~1 retains the smallest prefix whose cumulative normalized score reaches $\tau$, subject to the fixed bounds $K_{\min}=8$ and $K_{\max}=80$. A larger $\tau$ generally preserves more candidates and may improve coverage, but it also increases the candidate-pool size passed to Step~2.

\begin{table}[t]
\centering
\small
\caption{Sensitivity analysis of $\tau$ on the pooled KEV and ATCM validation partitions.}
\label{tab:tau_sensitivity}
\begin{tabular}{ccccc}
\toprule
$\tau$
& Label Cov.
& Full Cov.
& Partial Cov.
& Avg.\ Cand. \\
\midrule

0.50
& 0.8367
& 0.7015
& \textbf{0.9776}
& 75.66 \\

0.60
& \textbf{0.8453}
& \textbf{0.7164}
& \textbf{0.9776}
& 78.75 \\

0.70
& \textbf{0.8453}
& \textbf{0.7164}
& \textbf{0.9776}
& 79.64 \\

0.80
& \textbf{0.8453}
& \textbf{0.7164}
& \textbf{0.9776}
& 79.84 \\

$0.90^{\dagger}$
& \textbf{0.8453}
& \textbf{0.7164}
& \textbf{0.9776}
& 79.91 \\

\bottomrule
\end{tabular}

\par\vspace{3pt}
\footnotesize
$\dagger$ denotes the setting used in the main experiments.
\end{table}

Table~\ref{tab:tau_sensitivity} shows that increasing $\tau$ from 0.50 to 0.60 improves Label Coverage from 0.8367 to 0.8453 and Full Coverage from 0.7015 to 0.7164, while Partial Coverage remains at 0.9776. Beyond $\tau=0.60$, coverage saturates, and the average candidate-pool size increases only marginally from 78.75 to 79.91. We adopt $\tau=0.90$ as a conservative candidate-retention setting, prioritizing the preservation of potentially relevant techniques over aggressive early pruning. Since the candidate-pool size is already constrained by $K_{\max}=80$, a higher threshold introduces little additional computational overhead while allowing Step~2 to make the final ranking decisions based on richer CVE--technique interactions.


\subsubsection{Sensitivity to $\lambda_{\mathrm{freq}}$}

The frequency-penalty coefficient $\lambda_{\mathrm{freq}}$ appears in the final Step~2 scoring function and controls the penalty applied to frequently observed techniques.

\begin{table}[t]
\centering
\small
\caption{Sensitivity analysis of $\lambda_{\mathrm{freq}}$ on the KEV and ATCM validation partitions.}
\label{tab:lambda_freq_sensitivity}
\resizebox{\columnwidth}{!}{%
\begin{tabular}{c|ccc|ccc}
\toprule
& \multicolumn{3}{c|}{KEV}
& \multicolumn{3}{c}{ATCM} \\
$\lambda_{\mathrm{freq}}$
& NDCG@10 & MRR & Hit@10
& NDCG@10 & MRR & Hit@10 \\
\midrule

0.00
& 0.5875 & 0.7255 & \textbf{0.9661}
& \textbf{0.6238} & 0.6905 & 0.8933 \\

0.05
& \textbf{0.5930} & \textbf{0.7585} & \textbf{0.9661}
& 0.6051 & 0.6830 & 0.8933 \\

$0.10^{\dagger}$
& 0.5834 & 0.7151 & \textbf{0.9661}
& 0.6050 & 0.6853 & \textbf{0.9067} \\

0.20
& 0.5890 & 0.7469 & \textbf{0.9661}
& 0.5992 & \textbf{0.7018} & 0.8667 \\

\bottomrule
\end{tabular}%
}

\end{table}

Table~\ref{tab:lambda_freq_sensitivity} shows that the effect of $\lambda_{\mathrm{freq}}$ varies across datasets and ranking metrics. On KEV, Hit@10 remains unchanged at 0.9661 across all tested values, while NDCG@10 and MRR fluctuate. On ATCM, increasing the penalty from 0.00 to 0.10 improves Hit@10 from 0.8933 to 0.9067, whereas a stronger penalty of 0.20 reduces it to 0.8667. We adopt $\lambda_{\mathrm{freq}}=0.10$ as a moderate frequency regularization setting, intended to reduce the dominance of frequently observed techniques without excessively suppressing candidates supported by other evidence.

\subsubsection{Sensitivity to $\gamma$}

The margin $\gamma$ appears in the supervised pairwise loss used to train the Step~2 reranker. It controls the required score separation between ground-truth and incorrect candidates and is not used during inference.

\begin{table}[t]
\centering
\small
\caption{Sensitivity analysis of the pairwise margin $\gamma$ on the KEV and ATCM validation partitions.}
\label{tab:gamma_sensitivity}
\resizebox{\columnwidth}{!}{%
\begin{tabular}{c|ccc|ccc}
\toprule
& \multicolumn{3}{c|}{KEV}
& \multicolumn{3}{c}{ATCM} \\
$\gamma$
& NDCG@10 & MRR & Hit@10
& NDCG@10 & MRR & Hit@10 \\
\midrule

0.50
& 0.5819 & 0.7119 & 0.9492
& \textbf{0.6216} & 0.6914 & 0.8933 \\

$1.00^{\dagger}$
& 0.5834 & 0.7151 & \textbf{0.9661}
& 0.6050 & 0.6853 & \textbf{0.9067} \\

1.50
& \textbf{0.5921} & \textbf{0.7341} & \textbf{0.9661}
& 0.6095 & \textbf{0.6944} & 0.8800 \\

2.00
& 0.5859 & 0.6938 & \textbf{0.9661}
& 0.6078 & 0.6775 & 0.8933 \\

\bottomrule
\end{tabular}%
}

\end{table}

Table~\ref{tab:gamma_sensitivity} shows that ranking performance varies moderately across the tested margins, without a consistent improvement as $\gamma$ increases. On KEV, Hit@10 reaches 0.9661 for all margins greater than or equal to 1.00, while on ATCM, $\gamma=1.00$ achieves the highest Hit@10 of 0.9067. We adopt $\gamma=1.00$ as a standard unit-margin constraint, encouraging the reranker to distinguish relevant from irrelevant techniques without imposing an unnecessarily large score separation.

\section{CONCLUSION}
In this work, we formulate CVE-to-\A mapping as an evidence-constrained data integration and ranking problem. So we propose \S, a two-step framework that integrates heterogeneous security evidence to construct high-coverage candidate pools and prioritizes techniques through evidence-aware ranking. Experiments across six public datasets show that \S reduces the ATT\&CK search space by approximately 60\% while retaining at least one ground-truth technique for 95.11\%--100\% of CVEs.
The work offers a reusable design for integrating heterogeneous, incomplete, and inconsistent data sources to infer implicit relationships, with demonstrated effectiveness in vulnerability-to-adversary behavior mapping.
Given the growing volume and diversity of cybersecurity threat intelligence data, we hope this work inspires further research on evidence-aware data integration to support more reliable security analysis and data-driven decision-making.

\newpage

\bibliographystyle{ACM-Reference-Format}
\bibliography{reference}



\end{document}